\documentclass[aps,showpacs,nofootinbib,amssymb,preprint]{revtex4-1}
\usepackage{graphicx,epsfig}
\usepackage{simplewick}
\usepackage{slashed}
\usepackage[retainorgcmds]{IEEEtrantools}
\usepackage{color}
\usepackage{setspace}

\usepackage[pdftex]{hyperref}

\newcommand{\nn}{\nonumber}
\newcommand{\be}{\begin{equation}}
\newcommand{\ee}{\end{equation}}
\newcommand{\bea}{\begin{eqnarray}}
\newcommand{\eea}{\end{eqnarray}}

\def\siml{{\ \lower-1.2pt\vbox{\hbox{\rlap{$<$}\lower6pt\vbox{\hbox{$\sim$}}}}\ }}

\def\vz {\vec{z}}

\def\vw {\vec{w}}

\def\d{\delta}

\def\bdel{\bar{\partial}}

\def \beq  {\begin{equation}}
\def \eeq  {\end{equation}}
\def \beqar {\begin{eqnarray}}
\def \eeqar {\end{eqnarray}}

\begin{document}

\title{\boldmath 
The Yang-Mills vacuum wave functional in three dimensions at weak coupling
\unboldmath}
\author{Sebastian Krug and Antonio Pineda}
\affiliation{Grup de F\'\i sica Te\`orica, Universitat
Aut\`onoma de Barcelona, E-08193 Bellaterra, Barcelona, Spain}

\date{\today}

\begin{abstract}
\noindent
We compute the Yang-Mills vacuum wave functional in three dimensions at weak coupling with ${\cal O}(e^2)$ precision. We use two different methods to solve the functional Schroedinger equation. One of them generalizes to ${\cal O}(e^2)$ the method followed by 
Hatfield at ${\cal O}(e)$~\cite{Hatfield:1984dv}.  
The other uses the weak coupling version of the gauge invariant formulation of the Schroedinger equation and the ground state wave functional followed by Karabali, Nair, and Yelnikov \cite{Karabali:2009rg}. We compare both results and discuss the differences between them.

\end{abstract}
\maketitle

\section{Introduction}

The determination of the ground-state (or vacuum) wave functional of QCD, $\Psi[{\vec A}]$, is tantamount to solving QCD, as any observable (for instance the static potential or the spectrum of the theory) can then be obtained by the computation of the expectation value of the appropriate operator. Even if the exact solution is not known, properly chosen trial functions may give valuable information of the vacuum using variational methods (see for instance \cite{Kovner:2004rg}).

We are still far from obtaining the exact ground-state wave functional of QCD. Even obtaining approximated expressions is very complicated.
This is also true in the weak coupling limit. One reason is due to the requirement that the wave functional, in addition to satisfying the Schroedinger equation, has to be gauge invariant. This constraint is imposed by the Gauss law. Therefore, one can not use standard quantum mechanics perturbation theory in a straightforward manner. A procedure to overcome this problem was devised in the case of SU(2), and to ${\cal O}(e)$ in the weak coupling expansion, in Ref.~\cite{Hatfield:1984dv}. An alternative procedure has also been considered in Ref. \cite{Chan:1986bv} and worked out to the same order in $e$.
   
In this paper we are interested in the three dimensional version of QCD without light fermions (i.e. Yang-Mills theory or gluodynamics). The method outlined by Hatfield~\cite{Hatfield:1984dv} can also be applied to the three dimensional case and a general group SU(N) without major modifications. We do so in Sec. \ref{sec:Hatfield} and obtain the ${\cal O}(e)$ expression for a general group SU(N) in three dimensions. The result agrees with the expression obtained by 
transforming the four dimensional result of Ref.~\cite{Hatfield:1984dv} to the expected three dimensional counterpart.  The solutions obtained with this method satisfy the Schroedinger equation by construction but not necessarily the Gauss law, though it can be explicitly shown that it does at ${\cal O}(e)$. We then compute the ${\cal O}(e^2)$ wave functional in what is a completely new result. 
Again, this result satisfies the Schroedinger equation by construction but, at this order, it's not possible to explicitly check the Gauss law, due to the complexity of the resulting expressions. The resulting wave functional  is explicitly real (as expected for the ground-state functional) and we name it $\Psi_{GL}[{\vec A}]$, where $GL$ stands for the explicit use of the Gauss law.

The fact that gauge invariance can not be guaranteed in general is one important drawback of the previous method. The reason is that the Gauss law is only implemented partially for some terms in some intermediate expressions. Moreover, even this partial implementation of the Gauss law is difficult to automatize, as at each order it has to be tailored somewhat. 

One solution to the previous problem would be to reformulate the Schroedinger equation in terms of gauge invariant variables. One such formulation was originally worked out in Refs. \cite{Karabali:1995ps,Karabali:1996je,Karabali:1996iu,Karabali:1997wk,Karabali:1998yq} 
(for some introductory notes see \cite{Schulz:2000td}) and, more recently, in Ref. \cite{Karabali:2009rg}, where a modified approximation scheme was devised. The authors use a change of field variables, which become complex, to simplify the problem. Even though the original motivation of those works was to understand the strong coupling limit (the opposite limit we are considering in this paper) and confinement in three dimensions, it is not difficult to see that the approximation scheme worked out in Ref. \cite{Karabali:2009rg} could be easily reformulated to provide with a systematic expansion of the weak coupling limit. We do so in Sec. \ref{sec:KNY} of this paper and compute the ground-state wave functional to ${\cal O}(e^2)$. The vacuum wave functional is a function of the gauge invariant variables $J^a$, which we then transform to the original gauge variables ${\vec A}^a$. The resulting expression is gauge invariant by construction and also satisfies the Schroedinger equation by construction. We name it $\Psi_{GI}[{\vec A}]\equiv \Psi_{GI}[J({\vec A})]$, where $GI$ stands for the use of gauge invariant degree of freedom. However, the explicit expression has the very unpleasant feature of having a non-trivial imaginary term. 

We have then obtained two different expressions for the wave functionals: $\Psi_{GL}[{\vec A}]$ and $\Psi_{GI}[{\vec A}]$, which actually look completely different. We compare them in Sec. \ref{sec:comp}. At ${\cal O}(e)$ it is possible to show, after several manipulations and using the symmetries of the integrals, that they are equal (so at this order both of them are real and gauge invariant). Such brute force approach happens to be unfeasible 
at ${\cal O}(e^2)$ due to the complexity of the expressions. We need an organizing principle for the comparison. The approach we follow is to rewrite $\Psi_{GL}[{\vec A}]$ in terms of the gauge invariant variable $J^a$ and a gauge dependent field $\theta^a$. All $\theta^a$ dependent terms should vanish if $\Psi_{GL}[{\vec A}]$ is going to satisfy the Gauss law and we explicitly show that this happens. This means that both $\Psi_{GL}[{\vec A}]$ and $\Psi_{GI}[{\vec A}]$ are gauge invariant. We would then say that both should be equal, since both satisfy the Schroedinger equation. We actually find (after a rather lengthy computation) that both are almost but not completely equal. The difference is proportional to a bilinear real term. This is puzzling but there is a reason behind it: $\Psi_{GL}[{\vec A}]$ and $\Psi_{GI}[{\vec A}]$ satisfy "different" Schroedinger equations. $\Psi_{GL}[{\vec A}]$ 
was obtained using the unregulated Schroedinger equation, whereas $\Psi_{GI}[{\vec A}]$ was obtained after the 
Schroedinger equation in terms of $J^a$ variables was regularized. In this last case, regularization produces an extra term in the Schroedinger equation, producing in turn an extra term in the wave functional. Irrespectively of the above this comparison allows to rewrite $\Psi_{GI}[{\vec A}]$  in an explicitly real form. This is by far non-trivial, as the initial $\Psi_{GI}[J]$ was explicitly complex and dependent on complex variables. In particular there is a delicate cancellation between terms such that, after transforming this expression back to real variables, the wave function becomes real (actually in our comparison we work the other way around and transform $\Psi_{GL}[{\vec A}]$, which is real, in terms of the complex variables). This is an important test of several parts of the computation done in Ref. \cite{Karabali:2009rg}. 

We believe that the weak coupling reformulation of the approach followed in Ref. \cite{Karabali:2009rg} can be helpful to understand the meaning of the partial resummations performed in the approximation scheme used in this reference, though we do not explore this issue in this paper. Our ${\cal O}(e)$ or ${\cal O}(e^2)$ wave functional can also be used to test different trial functionals in the literature that claim to have the proper weak and strong coupling limit. Typically they reproduce the leading order weak coupling expansion but not the ${\cal O}(e)$ corrections. This is certainly the case with covariantization approaches where the exponent of the wave functional is approximated by a bilinear term in the $B$ fields (see for instance \cite{Greensite:2007ij,Greensite:2011pj}). Therefore, our results can hint to how those trial functions could be improved to correctly incorporate corrections in the weak coupling limit. 

\section{Determination of $\Psi_{GL}[{\vec A}]$}
\label{sec:Hatfield}

The Yang-Mills Lagrangian reads
\be
{\cal L}=-\frac{1}{4}G^{\mu\nu,a}G_{\mu\nu}^a
\,,
\ee
where 
\be
G_{\mu\nu}^a=\partial_{\mu}A_{\nu}^a-\partial_{\nu}A_{\mu}^a+ef^{abc}A^b_{\mu}A^c_{\nu}
\,,
\ee
$eG_{\mu\nu}=[D_{\mu},D_{\nu}]$, $D_{\mu}=\partial_{\mu} +eA_{\mu}$, $A_{\mu}=-iT^aA_{\mu}^a$, 
$G^{\mu\nu}=-iT^aG^{\mu\nu}_a$, $T^a$ are the SU(N) generators, and $[T^a,T^b]=if^{abc}T^c$. 

We will work in the Hamiltonian formalism and partially fix the gauge to $A_0=0$. Therefore, we work with the ${\vec A}=(A_1, A_2)$ components
 only and 
\be
{\vec D}={\vec \nabla} +e{\vec A} 
\,,
\ee
\be
B^a=\frac{1}{2}\epsilon_{jk}(\partial_jA_k-\partial_kA_j+e[A_j,A_k])^a
 = {\vec \nabla}\times\vec{A}^a +\frac{e}{2}f^{abc}\vec{A}^b\times\vec{A}^c
\,,
\ee
where ${\vec A}\times {\vec B} \equiv \epsilon_{ij}A_iB_j$, $\vec \nabla_i \equiv \partial_i=\partial/\partial x^i$ (
 for simplicity, we use the metric $\eta_{\mu\nu}=\mathrm{diag}(-1,+1,+1)$, so there is no sign difference between upper and lower spatial indices), and $B=-iT^aB^a$. 

In Ref. \cite{Hatfield:1984dv} the wave functional was computed to ${\cal O}(e)$ at weak coupling. It is possible to generalize the method used in this reference. We do so here and compute the ground state wave function to ${\cal O}(e^2)$. The ground state wave functional has to satisfy the Schroedinger equation\footnote{In the following we use the notation ($d=2$): $\int_x \equiv \int d^dx$, $\int_\slashed{k} \equiv \int \frac{d^2k}{(2\pi)^d}$, $\slashed{\delta}(\vec{k})\equiv (2\pi)^d\delta^{(d)}(\vec{k})$, and so on.} \footnote{Note that the ground state energy can be normalized to zero by moving it to the left-hand side of the equation and absorbing it in the $B^2$ term as a counterterm.}:
\be
\frac{1}{2}\int_x\left(-\frac{\delta}{\delta \vec{A}^a(\vec{x})} \cdot \frac{\delta}{\delta \vec{A}^a(\vec{x})} + B^a(\vec{x}) B^a(\vec{x}) \right) 
\Psi_{GL}[{\vec A}] = E \Psi_{GL}[{\vec A}]
\,,
\ee
and the Gauss law constraint
\be
(\vec{D}\cdot\vec{E})^a\Psi_{GL}[{\vec A}] =
i \left(
\vec \nabla \cdot \frac{\delta }{\delta {\vec A}_a}+ef^{abc}{\vec A}_b \cdot \frac{\delta }{\delta {\vec A}_c}
\right)\Psi_{GL}[{\vec A}]=0
\,.
\ee

Because we are talking of the ground state we expect the wave function to be real and have zero nodes. 
Therefore, it can be written as the exponential of a functional $F[\vec A]$ that does not diverge for finite ${\vec A}$:
\be
\Psi_{GL}[{\vec A}]=e^{-F_{GL}[{\vec A}]}=e^{-F_{GL}^{(0)}[{\vec A}]-eF_{GL}^{(1)}[{\vec A}]-e^2F_{GL}^{(2)}[{\vec A}]+{\cal O}(e^3)}
\,.
\ee
and satisfies the Gauss law
\be
\left(
\vec \nabla \cdot \frac{\delta }{\delta {\vec A}_a}+ef^{abc}{\vec A}_b \cdot \frac{\delta }{\delta {\vec A}_c}
\right)F_{GL}[{\vec A}]=0
\,.
\ee

\subsection{Order $e^0$}
$F_{GL}^{(0)}$ can be obtained in several ways. It is equivalent to solving the Schroedinger equation of the free theory with the free Gauss law, in other words, $N^2-1$ replicas of QED without light fermions. In order to solve these equations, it is convenient to rewrite them in momentum space using
\be
\label{FT}
{\vec A}(\vec x)=\int_\slashed{k} e^{i{\vec k}\cdot{\vec x}} {\vec A}(\vec k)
\,,
\qquad
\frac{\delta}{\delta \vec{A}^a(\vec{x})}
=\int_\slashed{k} e^{-i{\vec k}\cdot{\vec x}}
\frac{\delta}{\delta \vec{A}^a(\vec{k})}
\,.
\ee
We then have for the free-field Schroedinger equation
\be
\frac{1}{2}\int_\slashed{k} \left(-\frac{\delta}{\delta \vec{A}^a(\vec{k})} \cdot \frac{\delta}{\delta \vec{A}^a(-\vec{k})} + (\vec{k}\times\vec{A}^a(\vec{k}))(\vec{k}\times\vec{A}^a(-\vec{k})) \right) \Psi^{(0)}_{GL}[{\vec A}] = E^{(0)} \Psi_{GL}^{(0)}[{\vec A}]
\,,
\ee
giving the following equation for the leading order term of the wave functional exponent
\be 
\label{Sch_Free}
\int_\slashed{k} \frac{\delta F_{GL}^{(0)}[{\vec A}]}{\delta \vec{A}^a(\vec{k})} \cdot \frac{\delta F_{GL}^{(0)}[{\vec A}] }{\delta \vec{A}^a(-\vec{k})} = \int_\slashed{k} (\vec{k}\times\vec{A}^a(\vec{k})) (\vec{k}\times\vec{A}^a(-\vec{k}))
\,.
\ee
The free-field Gauss law reads:
\bea
\label{GL_free}
\vec{k}\cdot \frac{\delta F_{GL}^{(0)}[{\vec A}]}{\delta \vec{A}^a(\vec{k})} = 0
\,. 
\eea

Eq. (\ref{Sch_Free}) suggests $F_{GL}^{(0)}$ to be quadratic in ${\vec A}$: 
\be 
\label{F2g2}
F_{GL}^{(0)}[{\vec A}] = \int_\slashed{k}{A}_i^a(\vec{k}){A}_j^a(-\vec{k})g_{ij}({\vec k})
\,.
\ee
The tensor structure of $g_{ij}({\vec k})$ can be fixed by the free-field Gauss law, Eq. (\ref{GL_free}), which implies that 
$g_{ij}({\vec k})$ only depends on the transverse component of the momentum. Therefore
\be
g_{ij}({\vec k})=g({\vec k}){\cal P}_{ij}({\hat k})
\,,
\ee 
where ${\cal P}_{ij}=\delta_{ij}-k_ik_j/{\vec k}^2$ is the projector to the transverse component. We can now solve Eq. (\ref{Sch_Free}) 
and determine $g({\vec k})$. As the equation is quadratic there are two solutions, of which we take the one that leads to a normalizable wave functional, which is
\be 
\label{F2}
F_{GL}^{(0)}[{\vec A}] = \frac{1}{2}\int_\slashed{k}\frac{1}{E_k} (\vec{k}\times\vec{A}^a(\vec{k})) (\vec{k}\times\vec{A}^a(-\vec{k}))
\,,\label{FGL0}
\ee
where $E_k \equiv |\vec k|$. A detailed explanation of this derivation can be found in Sec. 10.2 (see also Sec. 11.2) of \cite{Hatfield:1992rz}. One can see that, even in the free-field case, the implementation of the Gauss law is not trivial.

\subsection{Order $e$}

At ${\cal O}(e)$ the Schroedinger equation splits into two equations (organized by powers of ${\vec A}$):
\be 
\label{OeF1}
\int_\slashed{k} \frac{\delta F_{GL}^{(0)}[{\vec A}]}{\delta \vec{A}^a(-\vec{k})} \cdot \frac{\delta F_{GL}^{(1)}[{\vec A}] }{\delta \vec{A}^a(\vec{k})} = \frac{i}{2}f^{abc}\int_{\slashed{k_1},\slashed{k_2},\slashed{k_3}} \slashed{\delta}\left(\sum_{i=1}^3 \vec{k}_i\right) (\vec{k}_1\times\vec{A}^a(\vec{k}_1)) (\vec{A}^b(\vec{k}_2)\times\vec{A}^c(\vec{k}_3))
\,,
\ee
\be 
\int_\slashed{k} \frac{\delta^2 F_{GL}^{(1)}[{\vec A}]}{\delta \vec{A}^a(-\vec{k})\delta \vec{A}^a(\vec{k})} = 
0
\,,
\ee
and the Gauss law constraint reads\footnote{Note that in $d=2$:
$
\vec{A}^c(-\vec{k}-\vec{p})\cdot \left(\vec{p}\times\left(\vec{p}\times\vec{A}^b(\vec{p})\right)\right)= -\left(\vec{p}\times\vec{A}^b(\vec{p})\right) (\vec{p}\times\vec{A}^c(-\vec{k}-\vec{p}))
$.\\ Other useful relations are  $({\vec k}\cdot {\vec A})({\vec k}\times {\vec B})-({\vec k}\times {\vec A})({\vec k}\cdot {\vec B})={\vec k}^2
({\vec A}\times {\vec B})$ and  $\epsilon_{ij}\epsilon_{kl}=\delta_{ik}\delta_{jl}-\delta_{il}\delta_{jk}$.}
\bea 
\vec{k}\cdot \frac{\delta F_{GL}^{(1)}[{\vec A}]}{\delta \vec{A}^a(\vec{k})} 
&=&
-i f^{abc}\int_{\slashed{p_1},\slashed{p_2}}\vec{A}^b(\vec{p}_1) \cdot \frac{\delta F_{GL}^{(0)}[{\vec A}]}{\delta \vec{A}^a(\vec{p}_2)} 
\slashed{\delta}({\vec p}_1-{\vec p}_2+{\vec k})
\nn\\
&=& 
-i f^{abc}\int_{\slashed{p}}\frac{1}{|\vec{p}|} (\vec{p}\times\vec{A}^b(-\vec{k}-\vec{p}))  \left(\vec{p}\times\vec{A}^c(\vec{p})\right) 
\,.
\eea

Using Eq.(\ref{FGL0}) the left-hand-side of Eq. (\ref{OeF1}) can be rewritten as follows:
\bea
&&\int_{\slashed{p}}\frac{1}{|\vec{p}|} (\vec{p}\times\vec{A}^a(\vec{p}))  \left(\vec{p}\times \frac{\delta F_{GL}^{(1)}[{\vec A}]}{\delta \vec{A}^a(\vec{p})}\right) \nn\\
&&\qquad =
\int_{\slashed{p}}\frac{1}{|\vec{p}|} 
\left\{
{\vec p}^2\left(\vec{A}^a({\vec p})\cdot \frac{\delta F_{GL}^{(1)}[{\vec A}]}{\delta \vec{A}^a(\vec{p})}\right) 
-
\left({\vec p}\cdot {\vec A}^a({\vec p})\right)
\left(\vec{p}\cdot \frac{\delta F_{GL}^{(1)}[{\vec A}]}{\delta \vec{A}^a(\vec{p})}\right) 
\right\}
\,,
\eea
where the second term of the right-hand-side is known because of the Gauss law.

We are now in the position to obtain $F^{(1)}$. We profit from the fact that the kernel can be taken to be completely symmetric\footnote{Any term antisymmetric in any of the two indices will vanish when multiplied by the gauge fields. This means that the kernel is not completely determined, as such terms can always be added.} under the interchange of any two fields $A_{i,a_i,x_i}$,  $A_{j,a_j,x_j}$. Therefore, the density of
$
\int_{\slashed{p}}|{\vec p}| \left(\vec{A}^a({\vec p})\cdot \frac{\delta F_{GL}^{(1)}[{\vec A}]}{\delta \vec{A}^a(\vec{p})}\right) 
$
can be related with the density of $ F_{GL}^{(1)}[{\vec A}]$. More specifically, if for a functional $F\left[\vec{A}^{a_1}({\vec k}_1),\ldots,\vec{A}^{a_n}({\vec k}_n)\right]$ of $n$ fields we have
\be
\int_{\slashed{p}}|{\vec p}| \left(\vec{A}^a({\vec p})\cdot \frac{\delta F[{\vec A}]}{\delta \vec{A}^a(\vec{p})}\right) = \int_{\slashed{k_1},\ldots,\slashed{k_n}}D\left[\vec{A}^{a_1}({\vec k}_1),\ldots,\vec{A}^{a_n}({\vec k}_n)\right]\,,
\ee
then
\be
 F[{\vec A}] = \int_{\slashed{k_1},\ldots,\slashed{k_n}}\frac{1}{|{\vec k}_1|+\ldots+|{\vec k}_n|}D\left[\vec{A}^{a_1}({\vec k}_1),\ldots,\vec{A}^{a_n}({\vec k}_n)\right]\,.
\ee
With this we finally obtain
\begin{IEEEeqnarray}{rCl}
\label{F1H}
F_{GL}^{(1)}[{\vec A}] &=&  i f^{abc} \int_{\slashed{k_1},\slashed{k_2},\slashed{k_3}}\slashed{\delta}\left(\sum_{i=1}^3 \vec{k}_i\right) \Bigg\{ \frac{1}{2(\sum_i^3|\vec{k}_i|)} (\vec{k}_1\times\vec{A}^a(\vec{k}_1)) (\vec{A}^b(\vec{k}_2)\times\vec{A}^c(\vec{k}_3))\nn\\
&& -\frac{1}{(\sum_i^3|\vec{k}_i|)|\vec{k}_1||\vec{k}_3|} (\vec{k}_1\cdot\vec{A}^a(\vec{k}_1)) (\vec{k}_3\times\vec{A}^b(\vec{k}_2)) (\vec{k}_3\times\vec{A}^c(\vec{k}_3)) \Bigg\} 
\,,
\end{IEEEeqnarray}
which is the three dimensional version of Hatfield's result (except for a different sign convention for $e$).

\subsection{Order $e^2$}

At ${\cal O}(e^2)$ the Schroedinger equation leads to the following equality
\be
\label{Sch_e2}
\frac{1}{2}\int_x
\left(
\frac{\delta^2 F_{GL}^{(2)}}{(\delta A_i^a)^2}-\frac{\delta F_{GL}^{(1)}}{(\delta A_i^a)}\frac{\delta F_{GL}^{(1)}}{(\delta A_i^a)}
-2\frac{\delta F_{GL}^{(0)}}{(\delta A_i^a)}\frac{\delta F_{GL}^{(2)}}{(\delta A_i^a)}
+\frac{1}{4}f^{abc}f^{ade}(\vec{A}^b\times\vec{A}^c)(\vec{A}^d\times\vec{A}^e)
\right)
=0
\,.
\ee
At this order $F^{(2)}_{GL}$ can have contributions with four, two and zero fields (there are no contributions with three or one field): 
$F^{(2)}_{GL}=F_{GL}^{(2,4)}+F_{GL}^{(2,2)}+F_{GL}^{(2,0)}$. There is no need to compute $F_{GL}^{(2,0)}$, as it just changes the normalization of the state, which we do not fix, or alternatively can be absorbed in a redefinition of the ground-state energy. Then, 
Eq. (\ref{Sch_e2}) can be split into two terms with two and four fields respectively:
\be
\label{Sch_e2_4}
\frac{1}{2}\int_x
\left(
-\frac{\delta F_{GL}^{(1)}}{(\delta A_i^a)}\frac{\delta F_{GL}^{(1)}}{(\delta A_i^a)}
-2\frac{\delta F_{GL}^{(0)}}{(\delta A_i^a)}\frac{\delta F_{GL}^{(2,4)}}{(\delta A_i^a)}
+\frac{1}{4}f^{abc}f^{ade}(\vec{A}^b\times\vec{A}^c)(\vec{A}^d\times\vec{A}^e)
\right)
=0
\,,
\ee
and
\be
\label{Sch_e2_2}
\frac{1}{2}\int_x
\left(
\frac{\delta^2 F_{GL}^{(2,4)}}{(\delta A_i^a)^2}
-2\frac{\delta F_{GL}^{(0)}}{(\delta A_i^a)}\frac{\delta F_{GL}^{(2,2)}}{(\delta A_i^a)}
\right)
=0
\,.
\ee
$F^{(0)}_{GL}$ and $F^{(1)}_{GL}$ have already been determined (see Eqs. (\ref{FGL0}) and (\ref{F1H})) and can be inserted into Eqs. (\ref{Sch_e2_4}) and (\ref{Sch_e2_2}), but we still have to implement the Gauss law, which at this order reads
\be
{\vec k} \cdot \frac{\delta F_{GL}^{(2,4)}}{\delta {\vec A}^a({\vec k})}=-if^{abc}\int_{\slashed{p_1},\slashed{p_2}}{\vec A}^b({\vec p}_1)
\cdot
\frac{\delta F_{GL}^{(1)}}{\delta A_i^a({\vec p}_2)}
\slashed{\delta}({\vec p}_1-{\vec p}_2+{\vec k})
\,,
\ee 
\be
{\vec k} \cdot \frac{\delta F_{GL}^{(2,2)}}{\delta {\vec A}^a({\vec k})}=0
\,.
\ee 
One first solves Eq. (\ref{Sch_e2_4}) and determines $F_{GL}^{(2,4)}$. Afterwards $F_{GL}^{(2,2)}$ is fixed by Eq. (\ref{Sch_e2_2}). In order to obtain $F_{GL}^{(2,4)}$ the procedure is similar to the one used for $F_{GL}^{(1)}$. The dependence on $F_{GL}^{(2,4)}$ is encoded in the 2nd term of Eq. (\ref{Sch_e2_4}), which we rewrite in the following way
\bea
&&
\int_{\slashed{p}}\frac{1}{|{\vec p}|} (\vec{p}\times\vec{A}^a(\vec{p}))  \left(\vec{p}\times \frac{\delta F_{GL}^{(2,4)}[{\vec A}]}{\delta \vec{A}^a(\vec{p})}\right) \nn\\
&&\qquad =
\int_{\slashed{p}}\frac{1}{|{\vec p}|} 
\left\{
{\vec p}^2\left(\vec{A}^a(\vec{p})\cdot \frac{\delta F_{GL}^{(2,4)}[{\vec A}]}{\delta \vec{A}^a(\vec{p})}\right) 
-
\left({\vec p}\cdot {\vec A}^a({\vec p})\right)
\left(\vec{p}\cdot \frac{\delta F_{GL}^{(2,4)}[{\vec A}]}{\delta \vec{A}^a(\vec{p})}\right) 
\right\}
\,.
\eea
Once again the second term on the right-hand-side is given by the Gauss law, which allows us to isolate $F_{GL}^{(2,4)}$. As above we use the fact that the kernel can be taken to be completely symmetric under the interchange of fields $A_{i,a_i,x_i}$, which lets us relate the density of
$
\int_{\slashed{p}}|{\vec p}| \left(\vec{A}^a({\vec p})\cdot \frac{\delta F_{GL}^{(2,4)}[{\vec A}]}{\delta \vec{A}^a(\vec{p})}\right) 
$
with the density of $F_{GL}^{(2,4)}[{\vec A}]$ and we finally obtain
\begin{IEEEeqnarray}{l}
\label{F4fromF3}
F_{GL}^{(2,4)} = -\frac{1}{2}\int_{\slashed{p},\slashed{k_1},\slashed{k_2},\slashed{q_1},\slashed{q_2}}\frac{1}{\sum^2_i(|\vec{k}_i|+|\vec{q}_i|)}\left(\frac{\delta F_{GL}^{(1)}}{\delta A^a_i(\vec{p})}\right)[\vec{k}_1,\vec{k}_2]\left(\frac{\delta F_{GL}^{(1)}}{\delta A^a_i(-\vec{p})}\right)[\vec{q}_1,\vec{q}_2] \nn\\
\qquad -if^{b_1b_2c}\int_{\slashed{p},\slashed{k_1},\slashed{k_2},\slashed{q_1},\slashed{q_2}}\frac{\slashed{\delta}(\vec{q}_1+\vec{q}_2-\vec{p})}{\sum_i(|\vec{k}_i|+|\vec{q}_i|)|\vec{q}_1|}(\vec{q}_1\cdot\vec{A}^{b_1}(\vec{q}_1))\left(\vec{A}^{b_2}(\vec{q}_2)\cdot\frac{\delta F_{GL}^{(1)}}{\delta \vec{A}^c(\vec{p})}[\vec{k}_1,\vec{k}_2]\right) \nn\\
\qquad +\frac{1}{8}f^{a_1a_2c}f^{b_1b_2c}\int_{\slashed{k_1},\slashed{k_2},\slashed{q_1},\slashed{q_2}}\frac{\slashed{\delta}(\sum_i(\vec{k}_i+\vec{q}_i))}{\sum_i(|\vec{k}_i|+|\vec{q}_i|)}(\vec{A}^{a_1}(\vec{k}_1)\times\vec{A}^{a_2}(\vec{k}_2))(\vec{A}^{b_1}(\vec{q}_1)\times\vec{A}^{b_2}(\vec{q}_2))
\,,
\end{IEEEeqnarray}
which explicitly reads
\begin{IEEEeqnarray}{l}
\label{F24GL}
F_{GL}^{(2,4)}=f^{abc}f^{cde}\int_{\slashed{k_1},\slashed{k_2},\slashed{q_1},\slashed{q_2}}
\slashed{\delta}\left(\sum_i (\vec{k}_i+\vec{q}_i)\right) \frac{1}{|\vec{k}_1|+|\vec{k}_2|+|\vec{q}_1|+|\vec{q}_2|} \Bigg\{  \nn\\
\quad \frac{1}{2(|\vec{k}_1|+|\vec{k}_2|+|\vec{k}_1+\vec{k}_2|)(|\vec{q}_1|+|\vec{q}_2|+|\vec{q}_1+\vec{q}_2|)} \Bigg\{\left(\vec{A}^d(\vec{q}_1)\times\vec{A}^e(\vec{q}_2)\right)
\Bigg[ -\frac{1}{4}|\vec{k}_1+\vec{k}_2|^2 \vec{A}^a(\vec{k}_1)\times\vec{A}^b(\vec{k}_2) \nn\\
\qquad\qquad +\frac{|\vec{k}_1+\vec{k}_2|}{|\vec{k}_2|}(\vec{k}_1+\vec{k}_2)\times\vec{A}^a(\vec{k}_1) (\vec{k}_2\cdot\vec{A}^b(\vec{k}_2)) +\frac{(\vec{k}_1+\vec{k}_2)\cdot\vec{k}_2}{|\vec{k}_1||\vec{k}_2|} (\vec{k}_1\cdot\vec{A}^a(\vec{k}_1)) (\vec{k}_2\times\vec{A}^b(\vec{k}_2)) \nn\\
\qquad\qquad +(\vec{k}_1\times\vec{A}^a(\vec{k}_1)) (\vec{k}_1+\vec{k}_2)\cdot\vec{A}^b(\vec{k}_2) \Bigg] \nn\\
\qquad +  (\vec{k}_1\times\vec{A}^a(\vec{k}_1)) (\vec{q}_1\times\vec{A}^d(\vec{q}_1)) \left(\vec{A}^b(\vec{k}_2)\cdot\vec{A}^e(\vec{q}_2)\right)\nn\\
\qquad + \frac{1}{|\vec{k}_1||\vec{k}_2|}\left[2\vec{k}_2\cdot\vec{A}^e(\vec{q}_2)-\frac{\vec{q}_1\cdot\vec{k}_2}{|\vec{q}_1||\vec{k}_2|}\vec{q}_2\cdot\vec{A}^e(\vec{q}_2)\right] (\vec{k}_1\cdot\vec{A}^a(\vec{k}_1)) (\vec{k}_2\times\vec{A}^b(\vec{k}_2)) (\vec{q}_1\times\vec{A}^d(\vec{q}_1)) \nn\\
\qquad + \frac{1}{|\vec{k}_1|} (\vec{k}_1\cdot\vec{A}^a(\vec{k}_1)) (\vec{k}_1+\vec{k}_2) \times\vec{A}^b(\vec{k}_2) \Bigg[  \frac{1}{|\vec{q}_2|} (\vec{q}_1+\vec{q}_2)\times\vec{A}^d(\vec{q}_1) (\vec{q}_2 \cdot\vec{A}^e(\vec{q}_2)) \nn\\
\qquad\qquad + \frac{2}{|\vec{q}_1+\vec{q}_2|} (\vec{q}_1\times\vec{A}^d(\vec{q}_1)) (\vec{q}_1+\vec{q}_2) \cdot\vec{A}^e(\vec{q}_2) \Bigg] \nn\\
\qquad -\frac{2(\vec{q}_1+\vec{q}_2)\cdot\vec{q}_1}{|\vec{k}_1+\vec{k}_2||\vec{k}_1| |\vec{q}_1||\vec{q}_2|}  (\vec{k}_1\cdot\vec{A}^a(\vec{k}_1)) (\vec{k}_1+\vec{k}_2) \times\vec{A}^b(\vec{k}_2) (\vec{q}_1\times\vec{A}^d(\vec{q}_1)) (\vec{q}_2 \cdot\vec{A}^e(\vec{q}_2)) \nn\\
\qquad +\frac{2\vec{k}_1\times\vec{k}_2}{|\vec{k}_1||\vec{k}_2| |\vec{q}_1+\vec{q}_2| |\vec{q}_2|}  (\vec{k}_1\cdot\vec{A}^a(\vec{k}_1)) (\vec{k}_2 \times\vec{A}^b(\vec{k}_2)) (\vec{q}_2\times\vec{A}^d(\vec{q}_1)) (\vec{q}_2 \times\vec{A}^e(\vec{q}_2)) \nn\\
\qquad +\frac{2}{|\vec{q}_1+\vec{q}_2||\vec{q}_2|}  (\vec{k}_1\times\vec{A}^a(\vec{k}_1)) (\vec{k}_1+\vec{k}_2) \times\vec{A}^b(\vec{k}_2) (\vec{q}_2\times\vec{A}^d(\vec{q}_1)) (\vec{q}_2 \times\vec{A}^e(\vec{q}_2)) \nn\\
\qquad -\frac{1}{|\vec{k}_2||\vec{q}_2|}  (\vec{k}_2\times\vec{A}^a(\vec{k}_1)) (\vec{k}_2 \times\vec{A}^b(\vec{k}_2)) (\vec{q}_2\times\vec{A}^d(\vec{q}_1)) (\vec{q}_2 \times\vec{A}^e(\vec{q}_2))
\Bigg\} \nn\\
\quad +\frac{1}{8} \left(\vec{A}^a(\vec{k}_1)\times\vec{A}^b(\vec{k}_2)\right) \left(\vec{A}^d(\vec{q}_1)\times\vec{A}^e(\vec{q}_2)\right) \nn\\
\quad + \frac{1}{|\vec{k}_1|(|\vec{q}_1|+|\vec{q}_2|+|\vec{q}_1+\vec{q}_2|)}  (\vec{k}_1\cdot\vec{A}^a(\vec{k}_1)) \Bigg\{ \frac{1}{2}  (\vec{k}_1+\vec{k}_2) \times\vec{A}^b(\vec{k}_2)  \left(\vec{A}^d(\vec{q}_1)\times\vec{A}^e(\vec{q}_2)\right) \nn\\
\quad\quad 
-  (\vec{q}_1 \times\vec{A}^d(\vec{q}_1)  \left(\vec{A}^b(\vec{k}_2)\times\vec{A}^e(\vec{q}_2)\right) -  \frac{1}{|\vec{q}_1+\vec{q}_2||\vec{q}_2|} (\vec{k}_1+\vec{k}_2) \times\vec{A}^b(\vec{k}_2) (\vec{q}_1+\vec{q}_2)\times\vec{A}^d(\vec{q}_1) (\vec{q}_2 \cdot\vec{A}^e(\vec{q}_2)) \nn\\
\quad \quad +  \frac{1}{|\vec{q}_1||\vec{q}_2|} (\vec{q}_2 \times\vec{A}^b(\vec{k}_2)) (\vec{q}_1\cdot\vec{A}^d(\vec{q}_1)) (\vec{q}_2 \times\vec{A}^e(\vec{q}_2)) \nn\\
\quad \quad  -  \frac{1}{|\vec{q}_1+\vec{q}_2||\vec{q}_2|} (\vec{k}_1+\vec{k}_2) \cdot\vec{A}^b(\vec{k}_2) (\vec{q}_2\times\vec{A}^d(\vec{q}_1)) (\vec{q}_2 \times\vec{A}^e(\vec{q}_2)) 
\Bigg\}
\Bigg\}
\,. 
\end{IEEEeqnarray}

Proceeding analogously for $F_{GL}^{(2,2)}$ we obtain
\be
\label{F22GLa}
F_{GL}^{(2,2)}=\frac{1}{2}\int_{\slashed{p},\slashed{k_1},\slashed{k_2}} \frac{1}{\sum_{i}^2|{\vec k}_i|}\slashed{\delta}(\vec{p}+\vec{k_1}+\vec{k_2})
\left(\frac{\delta^2 F_{GL}^{(2,4)}}{\delta A^a_i(\vec{p})\delta A^a_i(-\vec{p})}\right)[\vec{k}_1,\vec{k}_2]
\,.
\ee
A direct computation of this object turns out to be extremely cumbersome. We will need to wait until Sec. \ref{sec:comp}, where we will be able to relate $F^{(2,2)}_{GL}$ with a known term of $F^{(2,2)}_{GI}$. Its explicit expression in terms of the ${\vec A}$ fields can be found in Eq. (\ref{F2GLb}).

We have thus obtained the wave functional to ${\cal O}(e^2)$ by extending the method first devised in Ref. \cite{Hatfield:1984dv} to the next order. The different contributions to $\Psi_{GL}[{\vec A}]$ are summarized in Eqs.  (\ref{F2}), (\ref{F1H}), (\ref{F24GL}) and (\ref{F2GLb}).
This result satisfies the Schroedinger equation by construction. It is also explicitly real. On the other hand, we can not claim (a priori) that the Gauss law is satisfied, as it has only been used in some intermediate computations. At ${\cal O}(e)$ it is possible to directly check that the Gauss law is satisfied. A direct check at ${\cal O}(e^2)$ turns out to be extremely difficult to obtain, due to the complexity of the expressions involved. In Sec. \ref{sec:comp} we will devise a method to test the gauge invariance of the 
expression obtained in this section. Finally we want to stress that the computation we have performed in this section has been carried 
out without any regularization. The final result happens to be finite but formal manipulations have been performed on potentially divergent expressions. We will come back to this issue in Sec. \ref{sec:comp}.

\section{Determination of $\Psi_{GI}[\vec{A}]$}
\label{sec:KNY}

In the previous section we have been able to compute the ground-state wave functional at weak coupling at ${\cal O}(e^2)$. 
However, it is difficult to automatize the method. First, regularization issues have been completely skipped in the previous computation and, second, the Gauss law is implemented in a partial, and somewhat ad hoc, manner. This last problem could be overcome by reformulating the Schroedinger equation in terms of gauge invariant variables. One such formulation was originally worked out in Refs. 
\cite{Karabali:1995ps,Karabali:1996je,Karabali:1996iu,Karabali:1997wk,Karabali:1998yq}\footnote{In those references the regularization of the Schroedinger equation was also addressed, dealing then with the other potential problem of the computation of Sec. \ref{sec:Hatfield}.}.  Here we mainly follow Ref. \cite{Karabali:2009rg}, where a modified approximation scheme was devised.  Even though the original motivation of those works was to understand the strong coupling limit, it is not difficult to see that the approximation scheme worked out in Ref. \cite{Karabali:2009rg} could be reformulated to provide with a systematic expansion of the weak coupling limit. We do so here and compute the ground-state wave functional to ${\cal O}(e^2)$. The only relevant information for us will be the change of field variables used. The initial new field variables will be complex:
\be
\label{AbarA}
A:=\frac{1}{2}\left(A_1+iA_2\right) \,,\qquad \bar A:=\frac{1}{2}\left(A_1-iA_2\right)
\,.
\ee
Therefore, it is also convenient to change the space and momentum components to complex variables in the following way (note that $k$ and $z$ are defined with different signs):
\begin{IEEEeqnarray}{rClrCl}
z &=& x_1-ix_2, \qquad &\bar z &=& x_1+ix_2, \nn\\
k &=& \frac{1}{2}(k_1+ik_2),  &\bar k &=& \frac{1}{2}(k_1-ik_2), \quad {\vec k}\cdot {\vec x}=\bar k \bar z+kz, \\
\partial &=& \frac{1}{2}\left(\partial_1+i\partial_2\right), 
&\bar \partial &=& \frac{1}{2}\left(\partial_1-i\partial_2\right),  \quad \partial\bar\partial=\frac{1}{4}\vec\nabla^2\,.  \nn
\end{IEEEeqnarray}
$A$ and $\bar A$ are still gauge-dependent degrees of freedom. These were replaced by gauge invariant fields, named 
$J$, in Refs. \cite{Karabali:1995ps,Karabali:1996je,Karabali:1996iu,Karabali:1997wk,Karabali:1998yq}. We will then use the following change of variables: $(A_1,A_2) \rightarrow (A,\bar A) \rightarrow (J(A,\bar A), \bar A(A,\bar A))$, where the relation between both variables is the following:
\bea
\bar A^a&=&\bar A^a
\\
\nn
J^a&=&2i\left(M^{\dagger}\right)^{ac}A^c+\frac{2}{e}\left((\partial M^{\dagger})M^{\dagger-1}\right)^a=-\frac{1}{\bar\partial}{\vec \nabla}\times\vec{A}^a+
{\cal O}(e)
\,,
\eea
where $M^\dagger$ is an invertible matrix, which is a function of  $\bar A$, defined implicitly by the equation
\be
\bar A=\frac{1}{e}M^{\dagger-1}\bar\partial M^{\dagger}
\,,
\ee
which inverted yields (for a more compact expression see Eq. (5) of \cite{Karabali:1996iu})
\be
\label{Mexp}
M(x)=1-e\frac{4}{\vec \nabla^2}(\bar \partial A)+e^2\frac{4}{\vec \nabla^2}\bar \partial A\frac{4}{\vec \nabla^2}\bar \partial A
+{\cal O}(e^3)
\,,
\ee
\be
\label{Mdaggerexp}
M^{\dagger}(x)=1+e\frac{4}{\vec \nabla^2}( \partial \bar A)+e^2\frac{4}{\vec \nabla^2} \partial \left(\frac{4}{\vec \nabla^2} \partial\bar A\right) \bar A+{\cal O}(e^3)
\,.
\ee
These equalities naturally lead to consider the following Green functions:
\bea
\bar G(z) &\equiv& \frac{1}{\bar \partial_z}\delta^{(2)}(\vec z)=-i\int \frac{d^2k}{(2\pi)^2}e^{i\vec k \vec z}\frac{1}{\bar k}=\frac{1}{\pi}\frac{\bar z}{z\bar z+\epsilon^2}
\,,
\\
 G(z) &\equiv& \frac{1}{\partial_z}\delta^{(2)}(\vec z)=-i\int \frac{d^2k}{(2\pi)^2}e^{i\vec k \vec z}\frac{1}{k}=\frac{1}{\pi}\frac{z}{z\bar z+\epsilon^2}
\,.
\eea
Also, a useful relation reads 
%
\be
\frac{1}{\bar \partial} \left(\left(\frac{1}{\bar \partial} \bar A^a\right) \bar A^b\right)
=
-\frac{1}{\bar \partial} \left(\bar A^a \frac{1}{\bar \partial} \bar A^b\right)
+
\left(\frac{1}{\bar \partial} \bar A^a\right)
\left(\frac{1}{\bar \partial} \bar A^b\right)
\,,
\ee
which can easily be checked in momentum space.
We also need ($T_F=1/2$)
\bea
\left(M^{\dagger}\right)^{ac}=\frac{1}{T_F}Tr[T^aM^{\dagger}T^cM^{\dagger-1}]
\,.
\eea
The Gauss law operator can be written in a compact form in terms of $\bar A$ and $J$:
\be
I^a(\vec x)=({\vec D}\cdot {\vec E})^a(\vec x)
=
i\int_y\left(
D_x^{ab}\frac{\delta J^c(\vec y)}{\delta A^b(\vec x)}+{\bar D}_x^{ab}\frac{\delta J^c(\vec y)}{\delta \bar A^b(\vec x)}
\right)\frac{\delta }{\delta \bar J^c(\vec y)}
+i{\bar D}_x^{ab}\frac{\delta }{\delta \bar A^b(\vec x)}
\,.
\ee
Not surprisingly the dependence on $J$ drops out, since it is possible to prove that 
\be
 D_x^{ab}\frac{\delta J^c(\vec y)}{\delta A_i^b(\vec x)}+{\bar D}_x^{ab}\frac{\delta J^c(\vec y)}{\delta \bar A_i^b(\vec x)}=0
\,,
\ee 
where we have used the following properties (keep in mind that $M^{-1}_{ac}=M_{ca}$)
\bea
\frac{\delta J^c(\vec y)}{\delta A^b(\vec x)}&=&2i M^{\dagger}_{cb}(\vec y)\delta(\vec y-\vec x)
\,,
\\
\frac{\delta J^c(\vec y)}{\delta \bar A^b(\vec x)}&=&2\left[i\frac{\delta M^{\dagger}_{cd}(\vec y)}{\delta \bar A^b(\vec x)}A_d(\vec y)
+\frac{1}{e}\frac{\delta }{\delta \bar A^b(\vec x)}\left((\partial M^{\dagger}(\vec y))M^{\dagger-1}(\vec y)\right)_c
\right]
\,,
\\
\frac{\delta M^{\dagger}_{cd}(\vec y)}{\delta \bar A^b(\vec x)}&=&
e\left(\frac{1}{\bar D}\right)^{de}_{yx}(-f_{ebh})M_{hc}^{\dagger-1}(\vec x)=
e\left(\frac{1}{\bar D}\right)^{eb}_{yx} f_{edh} M_{hc}^{\dagger-1}(\vec y)
\,,
\\
\left(\frac{1}{\bar D}\right)^{de}_{yx}
&=&
\bar G(y-x)\left[M^{\dagger-1}(\vec y)M^{\dagger}(\vec x)\right]_{de}
\,.
\eea
Therefore we obtain
\be
I^a(\vec x)=i{\bar D}_x^{ab}\frac{\delta }{\delta \bar A^b(\vec x)}
\ee
for the Gauss law operator.

In Ref. \cite{Karabali:1998yq} it was shown that it was possible to write the Hamiltonian as a pure function of $J$ up to terms proportional to the Gauss law, which vanish when applied to physical (gauge-invariant) states. If we drop those terms the Hamiltonian reads\footnote{Note that in Ref. \cite{Karabali:1998yq} the normalization of $J$ is different.}
\bea
\label{HamiltonianNair}
H &=&  {2\over \pi} \int _{w,z} 
 {1\over (z-w)^2} {\d \over {\d J_a (\vw)}} {\d \over {\d
J_a (\vz)}} +{1\over 2} \int_z : \bdel J^a(\vec z) \bdel J^a(\vec z) :
\\
&&
+
 i e \int_{w,z} f_{abc} {J^c(w) \over \pi (z-w)} {\d \over {\d J_a (\vw)}} {\d \over {\d
J_b (\vz)}} 
+\frac{e^2C_A}{2\pi}  \int J_a (\vz) {\d \over {\d J_a (\vz)}} 
\,,
\nonumber
\eea
which we split into $H=H^{(0)}+H_I$, where $H^{(0)}$ is the first line 
and $H_I$ the second. It is important to note that the last term in Eq. (\ref{HamiltonianNair}) only appears after regularization of a divergent integral.

We can now obtain the vacuum wave functional in powers of $e$. We write 
\be
\label{WFJ}
\Psi_{GI}[J] = \exp (-F_{GI}[J])\,,
\ee 
where (following the notation of \cite{Karabali:2009rg})
\bea
-2F_{GI}[J] &=& \int f^{(2)}_{a_1 a_2}(\vec x_1, \vec x_2)\ J^{a_1}(\vec x_1) J^{a_2}(\vec x_2) ~+~
\frac{e}{2}\ f^{(3)}_{a_1 a_2 a_3}(\vec x_1,\vec  x_2,\vec  x_3)\ J^{a_1}(\vec x_1) J^{a_2}(\vec x_2) J^{a_3}(\vec x_3)
\nn\\
&&\hskip .2in~+~
\frac{e^2}{4}\ f^{(4)}_{a_1 a_2 a_3 a_4}(\vec x_1,\vec  x_2,\vec  x_3,\vec  x_4)\ J^{a_1}(\vec x_1) J^{a_2}(\vec x_2) J^{a_3}(\vec x_3)
J^{a_4}(\vec x_4)~+~\ldots\label{rec2}
\eea
and the kernels $f^{(2)}_{a_1 a_2}(\vec x_1, \vec x_2)$, 
$f^{(3)}_{a_1 a_2 a_3}(\vec x_1, \vec x_2, \vec x_3)$, {\it etc}., have the expansions
\bea
f^{(2)}_{a_1 a_2}(\vec x_1, \vec x_2) &=& f^{(2)}_{0~a_1 a_2}(\vec x_1, \vec x_2) +
e^2 f^{(2)}_{2~a_1 a_2}(\vec x_1, \vec x_2) +\ldots\nonumber\\
f^{(3)}_{a_1 a_2 a_3}(\vec x_1, \vec x_2, \vec x_3)&=& f^{(3)}_{0~a_1 a_2 a_3}(\vec x_1, \vec x_2, \vec x_3) + e^2 f^{(3)}_{2~a_1 a_2 a_3}(\vec x_1, \vec x_2, \vec x_3)
+\ldots\label{rec3}\\
f^{(4)}_{a_1 a_2 a_3 a_4}(\vec x_1, \vec x_2, \vec x_3, \vec x_4) &=& f^{(4)}_{0~a_1 a_2 a_3 a_4}(\vec x_1, \vec x_2, \vec x_3, \vec x_4) +\ldots
\,.
\nn 
\eea

Acting with the Hamiltonian of Eq. (\ref{HamiltonianNair}) onto this expansion of the wave functional and equating terms of equal numbers of $J$'s we obtain recursion relations for the kernels. These read
\beqar
 \label{rec4}
&& 2\frac{e^2C_A}{2\pi}~ f^{(2)}_{a_1 a_2}(\vec x_1, \vec x_2) + 4 \int_{x,y}  f^{(2)}_{a_1 a}(\vec x_1, \vec x) (\bar{\Omega}^0)_{ab}(\vec x,\vec y) f^{(2)}_{b a_2}(\vec y, \vec x_2) +V_{ab}  
\\
&&+e^2 \left[ 6 \int_{x,y} \!\! f^{(4)}_{a_1 a_2 a b }(\vec x_1, \vec x_2, \vec x,\vec y) (\bar{\Omega}^0)_{ab}(\vec x,\vec y) + 3 \int_{x,y} \!\! f^{(3)}_{a_1 a b  }(\vec x_1, \vec x,\vec y) (\bar{\Omega}^1)_{ab a_2}(\vec x,\vec y, \vec x_2)\right] 
= 0\nn
\eeqar
for the term with 2 $J$'s, while for the terms with $p \ge 3$ $J$'s the recursion relation is
\beqar
&&\frac{e^2C_A}{2\pi} p f^{(p)}_{a_1\cdots a_p} + \sum_{n=2}^{p} n(p+2-n) f^{(n)}_{a_1\cdots a_{n-1} a}(\bar{\Omega}^0)_{ab} f^{(p-n+2)}_{b a_n\cdots a_p} \nonumber \\
&&+ \sum_{n=2}^{p-1} n(p+1-n) f^{(n)}_{a_1\cdots a_{n-1}a} (\bar{\Omega}^1)_{ab a_p} f^{(p-n+1)}_{b a_n\cdots a_{p-1}} \nonumber\\
&&+ e^2 \left[ \frac{(p+1)(p+2)}{2}\ f^{(p+2)}_{a_1\cdots a_p a b}(\bar{\Omega}^0)_{ab} +\frac{p(p+1)}{2}\ f^{(p+1)}_{a_1\cdots a_{p-1} a b} (\bar{\Omega}^1)_{ab a_p}\right] =0 \label{rec5}
\,.
\eeqar
In these equations, we have used the abbreviations (following \cite{Karabali:2009rg})
\beqar
(\bar{\Omega}^0)_{ab}(\vec x,\vec y) &=& \delta_{ab} \partial_y \bar{G}(\vec x,\vec y) \nonumber 
\,,
\\
(\bar{\Omega}^1)_{abc}(\vec x,\vec y,\vec z) &=& -\frac{i}{2}\ f^{abc} \left[ \delta(\vec z-\vec y) + \delta(\vec z-\vec x)\right] \bar{G}(\vec x,\vec y) 
\,,
\nonumber \\
V_{ab}(\vec x,\vec y) &=& \delta_{ab} \int_z \bar{\partial}_z \delta(\vec z-\vec x) ~\bar{\partial}_z \delta(\vec z-\vec y) \label{rec6}
\,.
\eeqar
These equations are the same as the ones in Ref.~\cite{Karabali:2009rg} (which we have checked explicitly). Note that the splitting into $H^{(0)}$ and $H_I$ was different there, since the last term in Eq. (\ref{HamiltonianNair}) was included in $H^{(0)}$.

If one were able to solve the set of Eqs. (\ref{rec4}-\ref{rec5}) exactly, one would obtain the exact vacuum functional, without any truncation. Therefore, those equations are a perfect playground on which to try different resummation schemes (as it was done in Ref.~\cite{Karabali:2009rg}). Here we focus on the weak coupling expansion and solve those equations iteratively.

At the lowest (zeroth) order in $e$, we have to solve Eq. (\ref{rec4}) for 
$f^{(2)}_{0\ a_1 a_2}(\vec x_1, \vec x_2) $ with $e=0$. Note that this equation is quadratic in $f^{(2)}$, thus it has two solutions. We take the normalizable one, compatible with perturbation theory:

\be
\label{rec7}
f^{(2)}_{0\ a_1 a_2}(\vec x_1, \vec x_2)  = \delta_{a_1 a_2}
\frac{\bar \partial_{x_1}^2}{\sqrt{-\vec \nabla_{x_1}^2}}
\delta^{(2)}(\vec{x}_1- \vec{x}_2)
\Longleftrightarrow f^{(2)}_{0\ a_1 a_2}(\vec k) = -\frac{\bar{k}^2}{E_k} \delta_{a_1 a_2} \,,
\ee
where $E_k=|\vec k|$. 

At higher orders it is better to work in momentum space. We define
\bea
f^{(3)}_{a_1 a_2 a_3}(\vec x_1, \vec x_2, \vec x_3)&=& \int_{\slashed{k_1} \cdots \slashed{k_3}} \exp\left( i \sum_i^3 \vec k_i\cdot \vec x_i\right) \ f^{(3)}_{a_1 a_2 a_3}(\vec k_1, \vec k_2, \vec k_3)
\,, \\
f^{(4)}_{a_1 a_2 a_3 a_4}(\vec x_1, \vec x_2, \vec x_3, \vec x_4) &=& \int_{\slashed{k_1} \cdots \slashed{k_4}} \exp\left( i \sum_i^4 \vec k_i\cdot \vec x_i\right)\ f^{(4)}_{a_1 a_2 a_3 a_4}(\vec k_1, \vec k_2, \vec k_3, \vec k_4).
\label{rec8}
\eea
The recursive solution of equations (\ref{rec4}-\ref{rec5}) to order $e^2$ gives the following lowest order expressions for the cubic and quartic kernels:
\be
f^{(3)}_{0\ a_1 a_2 a_3}(\vec k_1, \vec k_2, \vec k_3) = -\frac{f^{a_1 a_2 a_3}}{24}\ (2\pi)^2 \delta (\vec k_1+\vec k_2+\vec k_3)\  g^{(3)}(\vec k_1,\vec k_2,\vec k_3)\label{rec50}
\,,
\ee
\be
f^{(4)}_{0\ a_1 a_2; b_1 b_2}(\vec k_1, \vec k_2; \vec q_1, \vec q_2) = \frac{f^{a_1 a_2 c} f^{b_1 b_2 c}}{64}\ (2\pi)^2 \delta (\vec k_1+\vec k_2+\vec q_1+\vec q_2)\ g^{(4)}(\vec k_1, \vec k_2; \vec q_1, \vec q_2) \label{rec51}
\,,
\ee
where
\beq \label{rec52}
g^{(3)}(\vec k_1,\vec k_2,\vec k_3) = \frac{16}{E_{k_1}\! + E_{k_2}\! + E_{k_3}}\left \{ \frac{\bar k_1 \bar k_2 (\bar k_1 - \bar k_2)}{E_{k_1} E_{k_2}} + {cycl.\ perm.} \right \}
\,,
\eeq
\begin{equation}\label{rec53}
\begin{array}{cl}
g^{(4)}(\vec k_1, \vec k_2; \vec q_1, \vec q_2)& =\ \vspace{.2in} \displaystyle \frac{1}{E_{k_1}\! + E_{k_2}\! + E_{q_1}\! + E_{q_2}} \\
\vspace{.2in}
&\!\!\!\!\displaystyle \left \{ g^{(3)}(\vec k_1, \vec k_2, -\vec k_1-\vec k_2)\ \frac{k_1 + k_2}{\bar k_1 +\bar k_2}\ g^{(3)}(\vec q_1, \vec q_2, -\vec q_1-\vec q_2) \right . \\
\vspace{.2in}
&\displaystyle -  \left [ \frac{(2\bar k_1 + \bar k_2)\,\bar k_1}{E_{k_1}} - \frac{(2\bar k_2 + \bar k_1)\,\bar k_2}{ E_{k_2}}\right ]\frac{4}{\bar k_1+\bar k_2}\  g^{(3)}(\vec q_1, \vec q_2, -\vec q_1-\vec q_2) \\
&\displaystyle -  \left .   g^{(3)}(\vec k_1, \vec k_2, -\vec k_1-\vec k_2)\ \frac{4}{\bar q_1+\bar q_2}\left [ \frac{(2\bar q_1 + \bar q_2)\,\bar q_1}{ E_{q_1}} - \frac{(2\bar q_2 + \bar q_1)\,\bar q_2}{ E_{q_2}}\right ] \right\} \,.
\\
\end{array}
\end{equation}

Note that the various $f^{(n)}$ are not fixed completely, since they are multiplied by local sources. Therefore, only the completely symmetric combination is determined, any antisymmetric term would vanish when multiplied by the sources, as they form a completely symmetric function.
 
Using the expressions for
 $f^{(3)}_0$, $f^{(4)}_0$ in Eq. (\ref{rec4}), the order $e^2$-term in $f^{(2)}$ is given by
\be
f_{2\ a_1 a_2}^{(2)}(\vec k) = \delta_{a_1 a_2}\frac{C_A}{2\pi}
{{\bar k}^2 \over E^2_k}
\left[ 
1 +
N
\right]
\label{f22Weak}
\,,
\ee
where
\be
N=\frac{E_k}{{\bar k}^2}\left(\int \frac{d^2 p}{32\pi}\ \frac{1}{\bar p}\ g^{(3)}(\vec k,\vec p,-\vec p-\vec k)\ +\ \int \frac{d^2 p}{64\pi}\ \frac{p}{\bar p}\ g^{(4)}(\vec k,\vec p;-\vec k,-\vec p)\right) \label{N}
\,.
\ee

It is possible to perform this integration, albeit numerically. 
The potentially divergent terms vanish after doing the integration over the phase of the complex number. We obtain
\be
N= 0.025999\,(8\pi) 
\,.
\ee
Note that it is real. This is not trivial to predict a priori since $g^{(3)/(4)}$ are complex functions. As we will see this is a strong check of the computation. The kernels $f^{(n)}$, $n\geq 5$, become nontrivial only at higher orders.

Note that the results above are nothing but Taylor expansions of the analogous set of Eqs. in Ref.~\cite{Karabali:2009rg} to the appropriate order. In practice this means setting $m=0$ in their computation and adding the first term in Eq. (\ref{f22Weak}). This 
last term will play a very important role in the comparison with the results of the previous section. 

Once we have an (approximated) expression for $\Psi_{GI}[J]$ we can transform it back to the 
original ${\vec A}$ variables: $\Psi_{GI}[J({\vec A})]\equiv \Psi_{GI}[{\vec A}]$. In principle, as it is a gauge invariant quantity, it should be possible to write it in terms of the gauge covariant quantities $\vec B$ and $\vec D$. However, since we work order by order in $e$, we do not need this. On the other hand, rotational $O(2)$ symmetry is preserved explicitly.\\
We will use the following relation to transform $J$ fields into $\vec{A}$ fields (where the derivatives are in the adjoint representation: $DB=\partial B+e[A,B]$; and we have defined $J=J^aT^a$):
\be
\bar \partial^nJ=-iM^{\dagger}(\bar D^{n-1}B)M^{\dagger-1}\,,
\ee
as well as Eqs. (\ref{Mexp}) and (\ref{Mdaggerexp}).

\subsection{Order $e^0$}

In this way at ${\cal O}(e^0)$ we obtain
\begin{IEEEeqnarray}{rCl}
 -2F^{(0)}_{GI}[{\vec A}]
 &=& -\int_\slashed{k}\frac{1}{E_k} (\vec{k}\times\vec{A}^a(\vec{k})) (\vec{k}\times\vec{A}^a(-\vec{k}))
 \,,
 \end{IEEEeqnarray}
which is the expected free-field expression.

\subsection{Order $e$}
 
At ${\cal O}(e)$ we obtain
\begin{IEEEeqnarray}{rCl}
\label{F1Nair}
F^{(1)}_{GI}[{\vec A}]
&=& i f^{abc} \int_{\slashed{k_1},\slashed{k_2},\slashed{k_3}}\slashed{\delta}\left(\sum_{i=1}^3 \vec{k}_i\right) \Bigg\{ \frac{1}{2|\vec{k}_1|} (\vec{k}_1\times\vec{A}^a(\vec{k}_1)) (\vec{A}^b(\vec{k}_2)\times\vec{A}^c(\vec{k}_3))\nn\\
&&- \frac{1}{|\vec{k}_3|\vec{k}_1^2} \left(  \frac{\vec{k}_1\times\vec{k}_2+i\vec{k}_1\cdot\vec{k}_2}{(|\vec{k}_1|+|\vec{k}_2|+|\vec{k}_3|)|\vec{k}_2|} +i  \right) (\vec{k}_1\times\vec{A}^a(\vec{k}_1))(\vec{k}_2\times\vec{A}^b(\vec{k}_2)) (\vec{k}_3\times\vec{A}^c(\vec{k}_3))\nn\\
&& + \frac{1}{|\vec{k}_3|\vec{k}_1^2} (\vec{k}_1\cdot\vec{A}^a(\vec{k}_1)) (\vec{k}_2\times\vec{A}^b(\vec{k}_2)) (\vec{k}_3\times\vec{A}^c(\vec{k}_3)) \Bigg\}
\,.
\end{IEEEeqnarray}
This term stems from a combination of $f^{(3)}$ and $f^{(2)}$ terms, as we have to remember that $J$ has an expansion in $e$ itself. 
Using the invariance of the integrals under interchange of integration variables and the fact that the delta function allows to write one momentum in terms of the other two, it is possible, however tedious and nontrivial, to show that the imaginary term of Eq. (\ref{F1Nair}) vanishes and that the real part is equal to Eq. (\ref{F1H}).

\subsection{Order $e^2$}

At ${\cal O}(e^2)$ we obtain
\begin{IEEEeqnarray}{rCl}
\label{F22GI}
 -2F^{(2,2)}_{GI}
 &=& \frac{C_A}{2\pi}\int_\slashed{k}\frac{1}{|\vec{k}|^2} (\vec{k}\times\vec{A}^a(\vec{k})) (\vec{k}\times\vec{A}^a(-\vec{k}))[1+N] 
\,.
\end{IEEEeqnarray}
This term is associated with the $f^{(2)}_2$ term.

 For the term with four gauge fields we obtain
\begin{IEEEeqnarray}{rCl}
&&
  -2\mathrm{Re}F^{(2,4)}_{GI}= 
  \\
  &&
  \frac{1}{4}f^{a_1a_2c}f^{b_1b_2c}\int_{\slashed{k_1},\slashed{k_2},\slashed{q_1},\slashed{q_2}} \slashed{\delta}(\vec{k_1}+\vec{k_2}+\vec{q_1}+\vec{q_2})\frac{1}{|\vec{k_1}+\vec{k_2}|} \left(\vec{A}^{a_1}(\vec{k_1})\times\vec{A}^{a_2}(\vec{k_2})\right) \left(\vec{A}^{b_1}(\vec{q_1})\times\vec{A}^{b_2}(\vec{q_2})\right)
\nn  \\
&& 
+ f^{a_1a_2c}f^{b_1b_2c}\int_{\slashed{k_1},\slashed{k_2},\slashed{q_1},\slashed{q_2}} \slashed{\delta}(\vec{k_1}+\vec{k_2}+\vec{q_1}+\vec{q_2})\frac{1}{\vec{k_2}^2} \left(\frac{1}{|\vec{k}_1+\vec{k_2}|} -\frac{1}{|\vec{k}_1|}\right) \nn\\
&&\qquad\qquad (\vec{k}_1\times\vec{A}^{a_1}(\vec{k}_1)(\vec{k}_2\cdot\vec{A}^{a_2}(\vec{k}_2))  (\vec{A}^{b_1}(\vec{q}_1)\times\vec{A}^{b_2}(\vec{q}_2)) 
\nn\\&& 
-f^{a_1a_2c}f^{b_1b_2c} \int_{\slashed{k_1},\slashed{k_2},\slashed{k_3},\slashed{k_4}}\slashed{\delta}\left(\sum_{i=1}^4\vec{k}_1\right) \Bigg\{\left(\frac{1}{|\vec{k}_1+\vec{k}_2|}-\frac{1}{|\vec{k}_3|}\right)\frac{1}{\vec{k_2}^2\vec{k_4}^2}(\vec{k}_1\times\vec{A}^{a_1}(\vec{k}_1))(\vec{k}_3\times\vec{A}^{b_1}(\vec{k}_3))\nn\\ 
  &&\quad\Bigg((\vec{k_2}\cdot\vec{A}^{a_2}(\vec{k_2}))(\vec{k}_4\cdot\vec{A}^{b_2}(\vec{k}_4)) - (\vec{k_2}\times\vec{A}^{a_2}(\vec{k_2})) (\vec{k}_4\times\vec{A}^{b_2}(\vec{k}_4)) \Bigg)\nn\\
  &&\quad + \frac{1}{|\vec{k}_2|(\vec{k}_3+\vec{k}_4)^2\vec{k}_3^2}(\vec{k}_1\times\vec{A}^{a_1}(\vec{k}_1)) (\vec{k}_2\times\vec{A}^{a_2}(\vec{k}_2)) \nn\\
  &&\quad \Bigg((\vec{k}_3\cdot\vec{A}^{b_1}(\vec{k}_3)) (\vec{k}_3+\vec{k}_4)\cdot\vec{A}^{b_2}(\vec{k}_4) - (\vec{k}_3\times\vec{A}^{b_1} (\vec{k}_3)) (\vec{k}_3+\vec{k}_4)\times\vec{A}^{b_2}(\vec{k}_4)\Bigg)\Bigg\}
\nn\\
&& 
+f^{a_1a_2c}f^{b_1b_2c}\int_{\slashed{k_1},\slashed{k_2},\slashed{k_3},\slashed{k_4}} \slashed{\delta}\left(\sum_{i=1}^4\vec{k}_i\right) \frac{\vec{k}_1\times\vec{k}_2}{(|\vec{k}_1|+|\vec{k}_2|+|\vec{k}_3+\vec{k}_4|)|\vec{k}_1||\vec{k}_2|} \nn\\
&&\qquad \Big(\frac{2}{|\vec{k}_3+\vec{k}_4||\vec{k}_1|} + \frac{1}{(\vec{k}_3+\vec{k}_4)^2} \Big) \nn\\
&&\qquad (\vec{k}_1\times\vec{A}^{a_1}(\vec{k}_1)) (\vec{k}_2\times\vec{A}^{a_2}(\vec{k}_2)) (\vec{A}^{b_1}(\vec{k}_3) \times \vec{A}^{b_2}(\vec{k}_4)) 
\nn\\
&& 
-2 f^{a_1a_2c}f^{b_1b_2c} \int_{\slashed{k_1},\slashed{k_2},\slashed{k_3},\slashed{k_4}} \slashed{\delta}\left(\sum_{i=1}^4\vec{k}_i\right)  (\vec{k}_1\times\vec{A}^{a_1}(\vec{k}_1) )(\vec{k}_2\times\vec{A}^{a_2}(\vec{k}_2) )(\vec{k}_3\times\vec{A}^{b_1}(\vec{k}_3))\nn\\
&&\quad \frac{1}{(|\vec{k}_1|+|\vec{k}_2|+|\vec{k}_3+\vec{k}_4|)|\vec{k}_1|}\Bigg( \frac{1}{|\vec{k}_3+\vec{k}_4|\vec{k}_2^2} \vec{k}_2\times \vec{A}^{b_2} (\vec{k}_4) \nn\\
&&\qquad  +\frac{1}{|\vec{k}_3+\vec{k}_4|\vec{k}_2^2 \vec{k}_4^2}\Big( \vec{k}_2 \times (\vec{k}_3-\vec{k}_1) (\vec{k}_4\cdot \vec{A}^{b_2} (\vec{k}_4)) + \vec{k}_2 \cdot (\vec{k}_3-\vec{k}_1) (\vec{k}_4\times \vec{A}^{b_2} (\vec{k}_4)) \Big)\nn\\
&&\qquad  -\frac{1}{|\vec{k}_2|\vec{k}_3^2 \vec{k}_4^2}  \Big(  \vec{k}_1 \times \vec{k}_3 (\vec{k}_4\cdot \vec{A}^{b_2} (\vec{k}_4)) - \vec{k}_1 \cdot \vec{k}_3 (\vec{k}_4\times \vec{A}^{b_2} (\vec{k}_4)) \Big)\nn\\
&&\qquad  + \frac{1}{|\vec{k}_2||\vec{k}_3+\vec{k}_4|^2 \vec{k}_3^2} \Big(\vec{k}_1 \times \vec{k}_3 (\vec{k}_3 + \vec{k}_4)\cdot \vec{A}^{b_2} (\vec{k}_4) - \vec{k}_1 \cdot \vec{k}_3 (\vec{k}_3 + \vec{k}_4)\times \vec{A}^{b_2} (\vec{k}_4)\Big)\Bigg)  
\nn\\&&
- f^{a_1a_2c}f^{b_1b_2c} \int_{\slashed{k_1},\slashed{k_2},\slashed{k_3},\slashed{k_4}} \slashed{\delta}\left(\sum_{i=1}^4\vec{k}_i\right) (\vec{k}_1\times\vec{A}^{a_1}(\vec{k}_1) )(\vec{k}_2\times\vec{A}^{a_2}(\vec{k}_2) )(\vec{k}_3\times\vec{A}^{b_1}(\vec{k}_3))(\vec{k}_4\times\vec{A}^{b_2}(\vec{k}_4)) \nn\\
&&
\quad \frac{1}{(\sum_i |\vec{k}_i|) (|\vec{k}_1|+|\vec{k}_2|+|\vec{k}_3+\vec{k}_4|)(|\vec{k}_3|+|\vec{k}_4|+|\vec{k}_1+\vec{k}_2|)|\vec{k}_1||\vec{k}_3|}  \nn\\
&&
\quad\Bigg\{  \frac{\vec{k}_1^2 \vec{q}_1^2-(\vec{k}_1\times\vec{k}_2)(\vec{q}_1\times\vec{q}_2)}{|\vec{k}_2||\vec{k}_4|(\vec{k}_1+\vec{k}_2)^2} 
\nn\\&&
\qquad - \frac{|\vec{k}_2|}{|\vec{k}_1+\vec{k}_2|} \left(2\left(2\frac{\vec{q}_1\cdot\vec{q}_2}{\vec{q}_2^2} +1\right) +4\frac{(\vec{k}_1\times\vec{k}_2)(\vec{q}_1\times\vec{q}_2)}{\vec{k}_2^2\vec{q}_2^2}\right) \left(1 - \frac{|\vec{k}_3|+|\vec{k}_4|+|\vec{k}_1+\vec{k}_2|}{|\vec{k}_1+\vec{k}_2|} \right) \nn\\\nn
&&
\qquad +  \left(\left(2\frac{\vec{k}_1\cdot\vec{k}_2}{\vec{k}_2^2} +1\right) \left(2\frac{\vec{q}_1\cdot\vec{q}_2}{\vec{q}_2^2} +1\right) -4\frac{(\vec{k}_1\times\vec{k}_2)(\vec{q}_1\times\vec{q}_2)}{\vec{k}_2^2\vec{q}_2^2}  \right) \left(1 - 2\frac{|\vec{k}_3|+|\vec{k}_4|+|\vec{k}_1+\vec{k}_2|}{|\vec{k}_1+\vec{k}_2|} \right) \Bigg\}   
\,,
\end{IEEEeqnarray}

\begin{IEEEeqnarray}{l}
 -2i\mathrm{Im}F^{(2,4)}_{GI} =\nn\\
i f^{a_1a_2c}f^{b_1b_2c} \int_{\slashed{k_1},\slashed{k_2},\slashed{k_3},\slashed{k_4}} \slashed{\delta}\left(\sum_{i=1}^4\vec{k}_i\right) (\vec{k}_1\times\vec{A}^{a_1}(\vec{k}_1)) (\vec{k}_2\times\vec{A}^{a_2}(\vec{k}_2)) (\vec{A}^{b_1}(\vec{k}_3) \times \vec{A}^{b_2}(\vec{k}_4))\nn\\
\quad \Bigg\{\frac{1}{(|\vec{k}_1|+|\vec{k}_2|+|\vec{k}_1+\vec{k}_2|)|\vec{k}_1||\vec{k}_2|} \left(\frac{\vec{k}_1^2+2\vec{k}_1\cdot\vec{k}_2}{|\vec{k}_1+\vec{k}_2||\vec{k}_1|} - \frac{\vec{k}_1^2+\vec{k}_1\cdot\vec{k}_2}{|\vec{k}_1+\vec{k}_2|^2} \right) - \frac{1}{\vec{k_2}^2} \left(\frac{1}{|\vec{k}_1+\vec{k_2}|} -\frac{1}{|\vec{k}_1|}\right) \Bigg\} \nn\\
+ if^{a_1a_2c}f^{b_1b_2c} \int_{\slashed{k_1},\slashed{k_2},\slashed{k_3},\slashed{k_4}} \slashed{\delta}\left(\sum_{i=1}^4\vec{k}_i\right) (\vec{k}_1\times\vec{A}^{a_1}(\vec{k}_1))(\vec{k_2}\times\vec{A}^{a_2}(\vec{k_2})) (\vec{k}_3\times\vec{A}^{b_1}(\vec{k}_3))(\vec{k}_4\cdot\vec{A}^{b_2}(\vec{k}_4)) \nn\\ 
\quad \Bigg\{\frac{1}{\vec{k}_1^2\vec{k}_2^2\vec{k}_3^2\vec{k}_4^2(\vec{k}_3+\vec{k}_4)^2} \Big( 2\vec{k}_1^2\vec{k}_3^2|\vec{k}_1+\vec{k}_2| -|\vec{k}_1|\vec{k}_3^2(\vec{k}_1+\vec{k}_2)^2 -\vec{k}_1^2|\vec{k}_3|(\vec{k}_1+\vec{k}_2)^2 + \vec{k}_1^2|\vec{k}_2|(\vec{k}_1+\vec{k}_2)^2 \nn\\ 
\qquad\quad + \vec{k}_1^2|\vec{k}_2|\vec{k}_3\cdot(\vec{k}_1+\vec{k}_2) \Big)    \nn\\
\qquad +2 \frac{1}{(|\vec{k}_1|+|\vec{k}_2|+|\vec{k}_3+\vec{k}_4|)|\vec{k}_1|} \Bigg\{\frac{\vec{k}_2 \cdot (2\vec{k}_1+\vec{k}_2)}{|\vec{k}_3+\vec{k}_4|\vec{k}_2^2 \vec{k}_4^2} -\frac{\vec{k}_1 \cdot \vec{k}_3}{|\vec{k}_2|\vec{k}_3^2 \vec{k}_4^2} + \frac{\vec{k}_1 \cdot \vec{k}_3}{|\vec{k}_2||\vec{k}_3+\vec{k}_4|^2 \vec{k}_3^2}\Bigg\} \Bigg\} \nn\\
+ if^{a_1a_2c}f^{b_1b_2c} \int_{\slashed{k_1},\slashed{k_2},\slashed{k_3},\slashed{k_4}} \slashed{\delta}\left(\sum_{i=1}^4\vec{k}_i\right) (\vec{k}_1\times\vec{A}^{a_1}(\vec{k}_1))(\vec{k_2}\times\vec{A}^{a_2}(\vec{k_2})) (\vec{k}_3\times\vec{A}^{b_1}(\vec{k}_3))(\vec{k}_3\cdot\vec{A}^{b_2}(\vec{k}_4)) \nn\\ 
\quad \Bigg\{ \frac{1}{|\vec{k}_2|(\vec{k}_3+\vec{k}_4)^2\vec{k}_3^2} -2 \frac{1}{(|\vec{k}_1|+|\vec{k}_2|+|\vec{k}_3+\vec{k}_4|)|\vec{k}_1|} \frac{1}{|\vec{k}_2||\vec{k}_3+\vec{k}_4|^2 \vec{k}_3^2} (\vec{k}_1 \cdot \vec{k}_3)\Bigg\} \nn\\
+ if^{a_1a_2c}f^{b_1b_2c} \int_{\slashed{k_1},\slashed{k_2},\slashed{k_3},\slashed{k_4}} \slashed{\delta}\left(\sum_{i=1}^4\vec{k}_i\right) (\vec{k}_1\times\vec{A}^{a_1}(\vec{k}_1))(\vec{k_2}\times\vec{A}^{a_2}(\vec{k_2})) (\vec{k}_3\cdot\vec{A}^{b_1}(\vec{k}_3)) (\vec{k}_3\times\vec{A}^{b_2}(\vec{k}_4)) \nn\\   
\qquad  \frac{1}{|\vec{k}_2|(\vec{k}_3+\vec{k}_4)^2\vec{k}_3^2} \nn\\
+ 2i f^{a_1a_2c}f^{b_1b_2c} \int_{\slashed{k_1},\slashed{k_2},\slashed{k_3},\slashed{k_4}} \slashed{\delta}\left(\sum_{i=1}^4\vec{k}_i\right)  (\vec{k}_1\times\vec{A}^{a_1}(\vec{k}_1) )(\vec{k}_2\times\vec{A}^{a_2}(\vec{k}_2) )(\vec{k}_3\times\vec{A}^{b_1}(\vec{k}_3)) (\vec{k}_3 \times \vec{A}^{b_2} (\vec{k}_4))\nn\\
\quad \frac{1}{(|\vec{k}_1|+|\vec{k}_2|+|\vec{k}_3+\vec{k}_4|)|\vec{k}_1|}\frac{1}{|\vec{k}_2||\vec{k}_3+\vec{k}_4|^2 \vec{k}_3^2} (\vec{k}_1 \times \vec{k}_3) \nn\\
+2i f^{a_1a_2c}f^{b_1b_2c} \int_{\slashed{k_1},\slashed{k_2},\slashed{k_3},\slashed{k_4}} \slashed{\delta}\left(\sum_{i=1}^4\vec{k}_i\right) (\vec{k}_1\times\vec{A}^{a_1}(\vec{k}_1) )(\vec{k}_2\times\vec{A}^{a_2}(\vec{k}_2) )(\vec{k}_3\times\vec{A}^{b_1}(\vec{k}_3))(\vec{k}_4\times\vec{A}^{b_2}(\vec{k}_4)) \nn\\
\quad \Bigg[\frac{2(\vec{k}_1\times\vec{k}_2)\Bigg\{|\vec{k}_2||\vec{k}_3| (\vec{k}_3\cdot\vec{k}_4)   - |\vec{k}_1| |\vec{k}_4|^3 +(\vec{k}_1+\vec{k}_2)^2 \left(2\vec{k}_3\cdot\vec{k}_4 +\vec{k}_4^2\right)   \Bigg\} }{(\sum_i |\vec{k}_i|) (|\vec{k}_1|+|\vec{k}_2|+|\vec{k}_3+\vec{k}_4|)(|\vec{k}_3|+|\vec{k}_4|+|\vec{k}_1+\vec{k}_2|)|\vec{k}_1|\vec{k}_2^2|\vec{k}_3|\vec{k}_4^2(\vec{k}_1+\vec{k}_2)^2}  
  \nn\\
\qquad +\frac{2(\vec{k}_1\times\vec{k}_2) \left(2\vec{k}_3\cdot\vec{k}_4 +\vec{k}_4^2\right) }{(|\vec{k}_1|+|\vec{k}_2|+|\vec{k}_3+\vec{k}_4|)(|\vec{k}_3|+|\vec{k}_4|+|\vec{k}_1+\vec{k}_2|)|\vec{k}_1|\vec{k}_2^2|\vec{k}_3|\vec{k}_4^2|\vec{k}_1+\vec{k}_2|} \nn\\
\qquad +\frac{1}{(|\vec{k}_1|+|\vec{k}_2|+|\vec{k}_3+\vec{k}_4|)|\vec{k}_1|} \Bigg\{\frac{-2 \vec{k}_2 \times \vec{k}_1}{|\vec{k}_3+\vec{k}_4|\vec{k}_2^2 \vec{k}_4^2} - \frac{\vec{k}_1 \times \vec{k}_3}{|\vec{k}_2|\vec{k}_3^2 \vec{k}_4^2} + \frac{ \vec{k}_1 \times \vec{k}_3}{|\vec{k}_2||\vec{k}_3+\vec{k}_4|^2 \vec{k}_3^2}  \Bigg\}   \Bigg]
\,.
\end{IEEEeqnarray}

The last two equations can be rewritten in several ways, yet, without an organizing principle, their sizes remain more or less the same. 

The resulting expression for the ground state wave functional seems to have a non-vanishing imaginary term. This is at odds with expectations, and with the result of the previous section. The real part does not look at all as the result obtained in the previous section either. We discuss this puzzling situation in the next section. 

\section{Comparison between both approaches}
\label{sec:comp}

If we compare the expressions we have found for the ground state wave functional in Secs. \ref{sec:Hatfield} and \ref{sec:KNY} we see that they look completely different. Even more so, whereas $\Psi_{GL}$ is explicitly real, $\Psi_{GI}$ has, a priori, a non-vanishing imaginary term. Only the ${\cal O}(e^0)$ expressions are trivially equal. 
Starting at ${\cal O}(e)$ we can get agreement between both expressions after quite lengthy and non-trivial rearrangements. 

At ${\cal O}(e^2)$ a direct comparison by brute force turns out to be completely impossible. In order to compare expressions we need an organizing principle to split the comparison into pieces. The procedure we follow is to rewrite $\Psi_{GL}$ in terms of $J$ and $\bar A$ (actually we will use the variable $\theta$ defined below\footnote{The field $\theta$ could be interpreted as a kind of generator of complex $SL(N,\mathbb{C})$ gauge transformations, see Ref. \cite{Karabali:1998yq}.}). If $\Psi_{GL}$ and $\Psi_{GI}$ are going to be equal, all terms proportional to $\bar A$ (or $\theta$) should vanish. Moreover, to a given order in $e$ the polynomial in $\bar A$ is finite so only a finite number of terms need to be compared. 

In order to perform this comparison to ${\cal O}(e^2)$ we need the following relations:
\begin{IEEEeqnarray}{rCl}
M^\dagger &\equiv& e^{e\theta}=1 + e\theta +\frac{e^2}{2}\theta^2 +{\cal O}(e^3) 
\,,\\
M^{\dagger -1} &=& 1 - e\theta +\frac{e^2}{2}\theta^2 +{\cal O}(e^3) 
\,,\\
A &=& -\frac{1}{2}M^{\dagger-1}J M^{\dagger} +\frac{1}{e}M^{\dagger-1}\partial M^{\dagger} \nn\\
&=& -\frac{1}{2}\left[J-e[\theta,J]+\frac{e^2}{4}\left[\theta,[\theta,J]\right]\right]
+\partial\theta   - \frac{e}{2}[\theta,\partial\theta] +\frac{e^2}{3!}\left[\theta,[\theta,\partial\theta]\right]+{\cal O}(e^3) 
\,,\\
\bar{A} &=& \frac{1}{e}M^{\dagger-1} \bar{\partial} M^{\dagger} 
= \bar{\partial}\theta - \frac{e}{2}[\theta,\bar \partial \theta] +\frac{e^2}{3!}\left[\theta,[\theta,\bar \partial\theta]\right]+{\cal O}(e^3) 
\,,\\
A^a(\vec{k}) &=& -\frac{i}{2}J^a(\vec{k}) +ik\theta^a(\vec{k})
+\frac{ie}{2}f^{abc}\int_\slashed{q}\theta^b(\vec{k}-\vec{q})J^c(\vec{q}) 
-\frac{ie}{2}f^{abc}\int_\slashed{q}q\,\theta^b(\vec{k}-\vec{q})\theta^c(\vec{q})
\nn\\
&& + \frac{ie^2}{4}f^{bcd}f^{dea}\int_\slashed{q}\int_\slashed{p} \theta^b(\vec{k}-\vec{q}-\vec{p}) J^c(\vec{q}) \theta^e(\vec{p})  
- \frac{ie^2}{3!}f^{bcd}f^{dea}\int_\slashed{q}\int_\slashed{p} \theta^b(\vec{k}-\vec{q}-\vec{p}) q\theta^c(\vec{q}) \theta^e(\vec{p}) 
\label{AinJ}\nn\\
&&
+{\cal O}(e^3) 
\,,\\
\bar{A}^a(\vec{k}) &=& i\bar{k}\theta^a(\vec{k}) 
-\frac{ie}{2}f^{abc}\int_\slashed{q}\bar{q}\,\theta^b(\vec{k}-\vec{q})\theta^c(\vec{q})
-\frac{ie^2}{3!}f^{bcd}f^{dea}\int_{\slashed{q},\slashed{p}}[k\bar{q}-\bar{k}q]\,\theta^b(\vec{k}-\vec{q}-\vec{p})\theta^c(\vec{q})\theta^e(\vec{p})
\label{AbarinJ}\nn\\
&&
+{\cal O}(e^3)
\,,
\end{IEEEeqnarray}
where $\theta = -i \theta^a T^a$, and we define the Fourier transform of $\theta$ and $J$ following the same conventions as in Eq. (\ref{FT}).

For the ${\cal O}(e^0)$ and the ${\cal O}(e)$ contributions of $F_{GL}$ it is possible to show that the $\theta$ terms vanish and the rest agrees with $F_{GI}$ in a direct fashion by just inserting the relations (\ref{AinJ}) and (\ref{AbarinJ}) into $F_{GL}^{(0)}$ and $F_{GL}^{(1)}$ and summing coefficients of terms with equal numbers of $J$'s and $\theta$'s. However, for the ${\cal O}(e^2)$ contributions, even after these simplifications, a brute force attack on the problem leads to expressions too large and complicated to directly show the equality of both expressions.\\
At this respect it is better to use some intermediate expressions of the $\Psi_{GL}$ computation that better agree with the structure of the $\Psi_{GI}$ result in terms of $J$. Particularly relevant for us is Eq. (\ref{F4fromF3}), which relates $F^{(2,4)}_{GL}$ with $(\delta F^{(1)}_{GL})/(\delta \vec{A})$. We can write $F^{(1)}_{GL}[J,\theta]\equiv F^{(1)}_{GL}[\vec{A}(J,\theta)]$ in terms of $g^{(3)}$. Using 
\begin{IEEEeqnarray}{rCl}
\label{dFdA}
\frac{\delta}{\delta A^a_i(\vec{p})} &=& \int_q\frac{\delta A^b(\vec{q})}{\delta A^a_i(\vec{p})}\frac{\delta}{\delta A^b(\vec{q})} 
+ \int_q\frac{\delta \bar{A}^b(\vec{q})}{\delta A^a_i(\vec{p})}\frac{\delta}{\delta \bar{A}^b(\vec{q})} \nn\\
&=& \int_{q_1,q_2}\frac{\delta A^b(\vec{q}_1)}{\delta A^a_i(\vec{p})}\frac{\delta J^c(\vec{q}_2)}{\delta A^b(\vec{q}_1)}\frac{\delta}{\delta J^c(\vec{q}_2)} + \int_{q_1,q_2}\frac{\delta \bar{A}^b(\vec{q}_1)}{\delta A^a_i(\vec{p})}\left(\frac{\delta J^c(\vec{q}_2)}{\delta \bar{A}^b(\vec{q}_1)}\frac{\delta}{\delta J^c(\vec{q}_2)}+\delta(\vec{q}_1-\vec{q}_2)\frac{\delta}{\delta \bar{A}^b(\vec{q}_2)}\right) \nn\\
&=&\frac{1}{2}\left(\delta_{1i}+i\delta_{2i}\right)(2i)\frac{\delta}{\delta J^a(\vec{p})}   + \frac{1}{2}\left(\delta_{1i}-i\delta_{2i}\right) \left(-2i\frac{p}{\bar{p}}\frac{\delta}{\delta J^a(\vec{p})}+\frac{\delta}{\delta \bar{A}^a(\vec{p})}\right)
+{\cal O}(e)  \label{funderiv}
\,,
\end{IEEEeqnarray}
we have
\begin{IEEEeqnarray}{rCl}
&&
\frac{\delta F^{(1)}_{GL}}{\delta A^a_i(\vec{p})}= -if^{aa_1a_2} \int_{\slashed{k_1},\slashed{k_2}}\slashed{\delta}\left(\vec{k}_1+\vec{k}_2+\vec{p}\right) \nn\\
&&
\quad
\Bigg\{\Bigg\{ (\delta_{1i}+i\delta_{2i})\frac{g^{(3)}(\vec{k}_1,\vec{k}_2,\vec{p})}{32} + (\delta_{1i}-i\delta_{2i})\left(-\frac{p}{\bar{p}}\frac{g^{(3)}(\vec{k}_1,\vec{k}_2,\vec{p})}{32}+\frac{\bar{k}_2^2}{2\bar{p}|\vec{k}_2|}\right)\Bigg\}J^{a_1}(\vec{k}_1)J^{a_2}(\vec{k}_2) \nn\\
&&
\quad
+\Bigg\{(\delta_{1i}+i\delta_{2i})\left( \frac{\bar{p}^2}{|\vec{p}|}-\frac{\bar{k}_1^2}{|\vec{k}_1|}\right)  - (\delta_{1i}-i\delta_{2i}) \left(\frac{1}{4}|\vec{p}|+\frac{\bar{k}_1}{|\vec{k}_1|}\left(-\frac{p}{\bar{p}}(\bar{k}_1+\bar{k}_2)+k_2\right) \right)\Bigg\}J^{a_1}(\vec{k}_1)\theta^{a_2}(\vec{k}_2) \nn\\
&&
\quad
+\Bigg\{(\delta_{1i}+i\delta_{2i})2\frac{\bar{p}k_1\bar{k}_2}{|\vec{p}|} -(\delta_{1i}-i\delta_{2i}) 2\frac{p k_1\bar{k}_2}{|\vec{p}|}\Bigg\} \theta^{a_1}(\vec{k}_1)\theta^{a_2}(\vec{k}_2) \Bigg\}+  O(e)\,.
\end{IEEEeqnarray}
With this we can write $F^{(2,4)}_{GL}[J,\theta]$ as a second order polynomial in $g^{(3)}$. This gives us the guiding principle to try to reconstruct $g^{(4)}$, which is also a second order polynomial in $g^{(3)}$. This term should be proportional to $J^4$ and we find that indeed it is.\\
In Eq. (\ref{F4fromF3}) one can see that all terms in $F^{(2,4)}_{GL}[J,\theta]$ have a prefactor of $\frac{1}{|\vec{k}_1|+|\vec{k}_2|+|\vec{q}_1|+|\vec{q}_2|}$. As we need the gauge ($\theta$) dependent terms to cancel with the corresponding terms from $F_{GL}^{(0)}$ and $F_{GL}^{(1)}$, that don't have this prefactor, we find a second guiding principle, which is to rewrite the $\theta$ dependent terms of $F^{(2,4)}_{GL}[J,\theta]$ in such a way, that this prefactor drops out and then try to find a form similar to the gauge dependent contributions of $F_{GL}^{(0)}$ and $F_{GL}^{(1)}$. To do so we extensively use the Jacobi identity and the invariance of the integrals under interchange of integration variables, as well as the delta function. We also use the fact that the integration kernels can be taken to be completely symmetric under the interchange of the variables of two equal fields (for instance $J^{a_1}(\vec{k}_1)J^{a_2}(\vec{k}_2)$). Still the computation is very lengthy and we will give some details in a different publication. In the end we obtain
\begin{IEEEeqnarray}{rCl}
\label{F0GLGI}
F^{(0)}_{GL} &=& \frac{1}{2}\int_\slashed{k}\frac{\bar{k}^2}{|\vec{k}|} J^a(\vec{k})J^a(-\vec{k}) 
+e\int_{\slashed{k_1},\slashed{k_2},\slashed{k_3}}\slashed{\delta}\left(\sum_{i=1}^3 \vec{k}_i\right) \frac{\bar{k_3}^2}{|\vec{k_3}|} f^{abc} J^a(\vec{k_1})\theta^b(\vec{k_2})J^c(\vec{k_3})\nn\\
&& -ef^{abc}\int_{\slashed{k_1},\slashed{k_2},\slashed{k_3}}\slashed{\delta}\left(\sum_{i=1}^3 \vec{k}_i\right) \frac{\bar{k}_1 (k_1\bar{k}_3-\bar{k}_1k_3)}{|\vec{k}_1|} J^a(\vec{k}_1)\theta^b(\vec{k}_2)\theta^c(\vec{k}_3) \nn\\
&& + \frac{e^2}{2}f^{a_1a_2c}f^{b_1b_2e}\int_{\slashed{k_1},\slashed{k_2},\slashed{q_1},\slashed{q_2}} 
\slashed{\delta}\left(\sum^2_i(\vec{k}_i+\vec{q}_i)\right) \left( \frac{(\bar{k}_1+\bar{k}_2)^2}{|\vec{k}_1+\vec{k}_2|} - \frac{\bar{k}_1^2}{|\vec{k}_1|} \right) J^{a_1}(\vec{k}_1) \theta^{a_2}(\vec{k}_2)J^{b_1}(\vec{q}_1)\theta^{b_2}(\vec{q}_2) 
\nn\\
&&+ e^2f^{a_1a_2c}f^{b_1b_2c}\int_{\slashed{k}_1,\slashed{k}_2,\slashed{q}_1,\slashed{q}_2}\slashed{\delta}\left(\sum_i(\vec{k}_i+\vec{q}_i)\right)  J^{a_1}(\vec{k}_1)\theta^{a_2}(\vec{k}_2) \theta^{b_1}(\vec{q}_1) \theta^{b_2}(\vec{q}_2)    \nn\\
&&\qquad\qquad \times \Bigg( \frac{1}{3} \frac{1}{|\vec{k}_1|} \bar{k}_1(k_1\bar{q}_2-\bar{k}_1q_2)  +\frac{1}{|\vec{q}_1+\vec{q}_2|}(\bar{q}_1+\bar{q}_2) (q_2\bar{q}_1-\bar{q}_2q_1)\, \Bigg)
 \nn\\
&& -2e^2f^{a_1a_2c}f^{b_1b_2c}\int_{\slashed{k_1},\slashed{k_2},\slashed{q_1},\slashed{q_2}} 
\slashed{\delta}\left(\sum_i(\vec{k}_i+\vec{q}_i)\right)  \theta^{a_1}(\vec{k}_1)\theta^{a_2}(\vec{k}_2)\theta^{b_1}(\vec{q}_1)  \theta^{b_2}(\vec{q}_2)    \frac{\bar{k_2}k_1\bar{q_2}q_1}{|\vec{k}_1+\vec{k}_2|} 
+O(e^3), 
\end{IEEEeqnarray}

\begin{IEEEeqnarray}{rCl}
\label{F1GLGI}
F^{(1)}_{GL} &=&  -f^{abc} \int_{\slashed{k_1},\slashed{k_2},\slashed{k_3}}\slashed{\delta}\left(\sum_{i=1}^3 \vec{k}_i\right) 
\frac{g^{(3)}(\vec{k}_1,\vec{k}_2,\vec{k}_3)}{96} 
 J^a(\vec{k}_1)J^b(\vec{k}_2)J^c(\vec{k}_3) 
 \nn
\\
&&
 -f^{abc} \int_{\slashed{k_1},\slashed{k_2},\slashed{k_3}}\slashed{\delta}\left(\sum_{i=1}^3 \vec{k}_i\right) 
\frac{\bar{k}_3^2}{|\vec{k}_3|} J^a(\vec{k}_1) \theta^b(\vec{k}_2)   J^c(\vec{k}_3) 
 \nn\\
&& -2f^{abc}\int_{\slashed{k_1},\slashed{k_2},\slashed{k_3}}\slashed{\delta}\left(\sum_{i=1}^3 \vec{k}_i\right)
\frac{\bar{k}_1k_2\bar{k}_3}{|\vec{k}_1|}J^a(\vec{k}_1)\theta^b(\vec{k}_2)\theta^c(\vec{k}_3) 
\nn\\
&& 
-ef^{a_1a_2c}f^{b_1b_2c}\int_{\slashed{k_1},\slashed{k_2},\slashed{q_1},\slashed{q_2}}\slashed{\delta}\left(\sum(\vec{k}_i+\vec{q}_i)\right) 
\frac{g^{(3)}(\vec{k}_1,\vec{k}_2,-\vec{k}_1-\vec{k}_2)}{32} J^{a_1}(\vec{k}_1)J^{a_2}(\vec{k}_2) J^{b_1}(\vec{q}_1) \theta^{b_2}(\vec{q}_2) \nn\\
&&
+ef^{a_1a_2c}f^{b_1b_2c} \int_{\slashed{k_1},\slashed{k_2},\slashed{q_1},\slashed{q_2}}  \slashed{\delta}\left(\sum_{i=1}^2 (\vec{k}_i+\vec
{q}_i)\right)   J^{a_1}(\vec{k_1})  J^{a_2}(\vec{k}_2) \theta^{b_1}(\vec{q}_1) \theta^{b_2}(\vec{q}_2)\nn\\
&&
\qquad\qquad  \times\Bigg(\frac{\bar{q}_2q_1}{\bar{q}_1+\bar{q}_2} \frac{g^{(3)}(\vec{k}_1,\vec{k}_2, -\vec{k}_1-\vec{k}_2)}{16} - \frac{\bar{q}_2}{(\bar{q}_1+\bar{q}_2)} \frac{\bar{k}_2^2}{2|\vec{k}_2|}\Bigg)  \nn\\
&&
 -ef^{a_1a_2c}f^{b_1b_2c} \int_{\slashed{k_1},\slashed{k_2},\slashed{q_1},\slashed{q_2}} 
\slashed{\delta}\left(\sum_{i=1}^2 (\vec{k}_i+\vec{q}_i)\right) 
\left( \frac{(\bar{k}_1+\bar{k}_2)^2}{|\vec{k}_1+\vec{k}_2|} -\frac{ \bar{k}_1^2}{|\vec{k}_1|} \right) 
J^{a_1}(\vec{k}_1) \theta^{a_2}(\vec{k}_2) J^{b_1}(\vec{q}_1)\theta^{b_2}(\vec{q}_2) \nn\\
&& 
-ef^{a_1a_2c}f^{b_1b_2c}\int_{\slashed{k_1},\slashed{k_2},\slashed{k_3},\slashed{q}}\slashed{\delta}\left(\sum_{i=1}(\vec{k}_i+\vec{q}_i)\right)    J^{a_1}(\vec{k}_1)\theta^{a_2}(\vec{k}_2) \theta^{b_1}(\vec{q}_1)\theta^{b_2}(\vec{q}_2)\nn\\
&&
\qquad\qquad \times\Bigg( \frac{\bar{k_1}}{|\vec{k}_1|} (k_1\bar{q}_2-\bar{k}_1q_2) +4\frac{(\bar{k_1}+\bar{k_2})}{|\vec{k}_1+\vec{k}_2|}q_1\bar{q_2}\Bigg) \nn \\
&&
+4ef^{a_1a_2c}f^{b_1b_2c}\int_{\slashed{k_1},\slashed{k_2},\slashed{q_1},\slashed{q_2}} \slashed{\delta}\left(\sum^2_i(\vec{k}_i+\vec{q}_i)\right)  \theta^{a_1}(\vec{k}_1)\theta^{a_2}(\vec{k}_2)\theta^{b_1}(\vec{q}_1)  \theta^{b_2}(\vec{q}_2)  \frac{k_1\bar{k_2}q_1\bar{q_2}}{|\vec{q}_1+\vec{q}_2|} +O(e^2),
\end{IEEEeqnarray}

\begin{IEEEeqnarray}{rCl}
\label{F24GLGI}
F^{(2,4)}_{GL} &=& -\frac{1}{512}f^{a_1a_2c}f^{b_1b_2c} \int_{\slashed{k_1},\slashed{k_2},\slashed{q_1},\slashed{q_2}}
\slashed{\delta}\left(\sum^2_i(\vec{k}_i+\vec{q}_i)\right) g^{(4)}(\vec k_1,\vec k_2;\vec q_1,\vec q_2) J^{a_1}(\vec{k}_1)J^{a_2}(\vec{k}_2)J^{b_1}(\vec{q}_1)J^{b_2}(\vec{q}_2) \nn\\
&& +\frac{1}{32}f^{a_1a_2c}f^{b_1b_2c}\int_{\slashed{k_1},\slashed{k_2},\slashed{q_1},\slashed{q_2}} 
\slashed{\delta}\left(\sum^2_i(\vec{k}_i+\vec{q}_i)\right) g^{(3)}(\vec{k}_1,\vec{k}_2,-\vec{k}_1-\vec{k}_2) J^{a_1}(\vec{k}_1)J^{a_2}(\vec{k}_2)J^{b_1}(\vec{q}_1)\theta^{b_2}(\vec{q}_2)\nn\\
&& + \frac{1}{2} f^{a_1a_2c}f^{b_1b_2c} \int_{\slashed{k_1},\slashed{k_2},\slashed{q_1},\slashed{q_2}} 
\slashed{\delta}\left(\sum^2_i(\vec{k}_i+\vec{q}_i)\right)   \left( \frac{(\bar{k}_1+\bar{k}_2)^2}{|\vec{k}_1+\vec{k}_2|} -  \frac{\bar{k}_1^2}{|\vec{k}_1|} \right) J^{a_1}(\vec{k}_1)\theta^{a_2}(\vec{k}_2)J^{b_1}(\vec{q}_1)\theta^{b_2}(\vec{q}_2) \nn\\
&& - f^{a_1a_2c}f^{b_1b_2c} \int_{\slashed{k_1},\slashed{k_2},\slashed{q_1},\slashed{q_2}} \slashed{\delta}
\left(\sum^2_i(\vec{k}_i+\vec{q}_i)\right)    J^{a_1}(\vec{k}_1)J^{a_2}(\vec{k}_2)\theta^{b_1}(\vec{q}_1)\theta^{b_2}(\vec{q}_2) \nn\\
&& \qquad\qquad \times \left(\frac{q_1\bar{q}_2}{\bar{q}_1+\bar{q}_2} \frac{g^{(3)}(\vec{k}_1,\vec{k}_2, -\vec{k}_1-\vec{k}_2)}{16} -   \frac{\bar{q}_2}{\bar{q}_1+\bar{q}_2} \frac{\bar{k}_2^2}{2|\vec{k}_2|}\right) \nn\\
&& + 2 f^{a_1a_2c}f^{b_1b_2c}\int_{\slashed{p},\slashed{k_1},\slashed{k_2},\slashed{q_1},\slashed{q_2}} 
\slashed{\delta}\left(\sum_i(\vec{k}_i+\vec{q}_i)\right) q_1\bar{q}_2 \left(\frac{\bar{k}_1+\bar{k}_2}{|\vec{k}_1+\vec{k}_2|} - \frac{\bar{k}_1}{|\vec{k}_1|} \right) 
J^{a_1}(\vec{k}_1)\theta^{a_2}(\vec{k}_2) \theta^{b_1}(\vec{q}_1)\theta^{b_2}(\vec{q}_2) 
\nn\\
&&  -2 f^{a_1a_2c}f^{b_1b_2c}\int_{\slashed{k_1},\slashed{k_2},\slashed{q_1},\slashed{q_2}} \slashed{\delta}\left(\sum_i(\vec{k}_i+\vec{q}_i)\right) \frac{ k_1\bar{k}_2q_1\bar{q}_2}{|\vec{k}_1+\vec{k}_2|} \theta^{a_1}(\vec{k}_1)\theta^{a_2}(\vec{k}_2)\theta^{b_1}(\vec{q}_1)  \theta^{b_2}(\vec{q}_2) +O(e)\,.
\end{IEEEeqnarray}

We now move to $F^{(2,2)}_{GL}$, which is associated to a one-loop computation. We have already mentioned in Sec. \ref{sec:Hatfield} that its direct determination in terms of ${\vec A}$ fields is not feasible. Again, we follow the strategy of rewriting $F^{(2,2)}_{GL}$ in terms of $J$ and $\theta$. For this we use Eq. (\ref{F24GLGI}), which we plug into Eq. (\ref{F22GLa}) after having rewritten the functional derivatives in terms of $J$ and $\bar A$ using Eq. (\ref{funderiv}). The calculation simplifies a lot and we find
\begin{IEEEeqnarray}{rCl}
F^{(2,2)}_{GL}  &=& - \frac{c_A}{32}\int_{\slashed{p},\slashed{k}}\frac{1}{|\vec{k}|}\left( \frac{1}{\bar{p}}  g^{(3)}(\vec k,\vec p,-\vec k-\vec p) + \frac{1}{2} \frac{p}{\bar{p}}  g^{(4)}(\vec p,\vec k;-\vec p,-\vec k) \right) J^{a}(\vec{k})J^{a}(-\vec{k}) +{\cal O}(e)\,.
\nn\\
\end{IEEEeqnarray}
This result allows us to write $F^{(2,2)}_{GL}$ in terms of the gauge fields. It reads
\be
\label{F2GLb}
F^{(2,2)}_{GL}  =-N\frac{C_A}{4\pi}\int_\slashed{k}\frac{1}{|\vec{k}|^2} (\vec{k}\times\vec{A}^a(\vec{k})) (\vec{k}\times\vec{A}^a(-\vec{k})) 
\,,
\ee
where $N$ has been defined in Eq. (\ref{N}).

We can now combine all the different contributions (in an, again, not completely trivial computation). We obtain the following equalities
\be
F_{GL}[\vec A(J,\theta)]=F_{GI}[J]+\frac{C_Ae^2}{4\pi}\int_\slashed{k}\frac{\bar{k}^2}{|\vec{k}|^2} J^a(\vec{k})J^a(-\vec{k}) +{\cal O}(e^3)
\,,
\ee
or in terms of the gauge fields
\be
F_{GI}[J({\vec A})]=F_{GL}[\vec A]-\frac{C_Ae^2}{4\pi}\int_\slashed{k}\frac{1}{|\vec{k}|^2} (\vec{k}\times\vec{A}^a(\vec{k})) (\vec{k}\times\vec{A}^a(-\vec{k})) +{\cal O}(e^3)
\,.
\ee
The first equality implies that $F_{GL}[\vec A]$ is gauge invariant to ${\cal O}(e^2)$, the second that $F_{GI}[J]$ is real to ${\cal O}(e^2)$.
We stress that $F^{(0)}_{GL}$, $F^{(1)}_{GL}$, and $F^{(2,4)}_{GL}$ are real, which is not evident at all as written in Eqs. (\ref{F0GLGI}), (\ref{F1GLGI}) and (\ref{F24GLGI}).

Overall we get complete agreement except for one bilinear real extra term in $F_{GI}$. Its origin can be traced back to the appearance of the last term of the Schroedinger equation in Eq. (\ref{HamiltonianNair}). In turn this term appears from an anomaly-like computation only after the kinetic operator has been regularized. Note that $F_{GL}$ was obtained without regularizing the theory, working with formal expressions.  The existence of very lengthy and complicated expressions in the intermediate steps impedes in practice the identification of the divergences. We expect these divergences to particularly affect $F_{GL}^{(2,2)}$, since we have functional derivatives acting on the wave functional density (see Eq.~(\ref{F22GLa})) that effectively produce contractions of fields and internal integrals over momenta.
Therefore, one could miss some contributions (and yet get a finite result) if formally manipulating the integrals as if they were finite before regulating them.
For the other terms of $F$ we have got a double check, which gives us strong confidence in our result. 

\section{Conclusions}

We have computed the Yang-Mills vacuum wave functional in three dimensions at weak coupling with ${\cal O}(e^2)$ precision. We have used two different methods to solve the Schroedinger functional equation: 
(A) One of them generalizes to ${\cal O}(e^2)$ the method followed by Hatfield at ${\cal O}(e)$~\cite{Hatfield:1984dv}. We have named the 
result $\Psi_{GL}[{\vec A}]$. 
(B) The other uses the weak coupling version of the gauge invariant formulation of the Schroedinger equation and the ground-state wave functional followed by Karabali, Nair, and Yelnikov \cite{Karabali:2009rg}. We have named the result $\Psi_{GI}[J]$. Each method has its own strengths and weaknesses, and they are to some extent complementary.

The computations performed with method (A) are relatively simple and the results are explicitly real. The generalization to four 
dimensions of the ${\cal O}(e^2)$ computation does not present major conceptual problems. Note that this is the order at which we expect to start to see the running of the coupling constant in $D=4$. On the other hand, such computation has two major drawbacks. First, the implementation of the Gauss law is not done in a systematic way, only partially in some intermediate steps. Therefore, we cannot guarantee a priori that the final result is gauge invariant. As the results rapidly grow in size and complexity, a direct check turns out to be unfeasible. Actually we were only able to check the Gauss law with the help of method (B). The major drawback, however, is that the computation has been performed with an unregulated kinetic operator. Whereas all computations can formally be carried out obtaining a finite result, some terms may be missed in this way. 

The computations with method (B) are somewhat more involved. Rather lengthy expressions appear when we rewrite the wave functional in terms of the gauge fields ${\vec A}$, which, moreover, look complex. Trying to prove by brute force that the result is real turns out to be impossible. Actually, we only manage to prove it after a careful comparison with the result of method (A). 
Moreover, a possible generalization to four dimensions does not look trivial. 
On the other hand, method (B) is particularly appealing, as it directly works with gauge-invariant degrees of freedom. Therefore, the 
Gauss law is automatically satisfied and it is not necessary to explicitly impose this constraint. Note also that the set of Eqs. (\ref{rec4}) and (\ref{rec5}) can be solved recursively. Therefore, it could be possible to automatize the computation and obtain the wave functionals at higher orders with a combination of algebraic/numeric programing. Finally, and most important, the kinetic operator had been regularized. This produced nontrivial 
contributions. 

We have compared both results. It is impossible to show that they are equal in a direct way. The strategy we follow helps a lot, yet it continues to be extremely complicated to prove the equality of the two expressions.  As we have already mentioned, this comparison has allowed us on the one hand to prove that $\Psi_{GL}$ is indeed gauge invariant and on the other hand that $\Psi_{GI}$ is real. Most interestingly, the agreement between both results is almost complete except for one extra term that appears with method (B). This term shows up from an anomaly-like computation once the theory is regularized. Such a contribution does not show up in method (A). Apparently, this is due to the fact that no regularization was used in this computation. This result is potentially very interesting because it is precisely this term that produces the mass gap 
and a linearly rising potential in the strong coupling limit in Ref. \cite{Karabali:1998yq}. Therefore, it's important to understand how (and if) such a term can be generated in a regulated version of the Schroedinger formalism in terms of the gauge fields, as this contribution has not been checked with an independent method so far. However, as regularization in the Schroedinger formalism with gauge variables is, to a large extent, uncharted territory, this requires a dedicated study beyond the aim of this work. We plan to address this issue in the near future, as well as to revisit the regularization with method (B), with the aim of resolving the discrepancy between the wave functions that we have found in this paper. In this context, it may be worth mentioning that supersymmetric extensions of Yang-Mills theory with ${\cal N} \geq 2$ do not have this term~\cite{Agarwal:2012bn}. This is not completely unexpected, as the introduction of supersymmetry improves the ultraviolet behavior of the theory. This may lead to convergent integrals and the disappearance of this extra term. Finally, we expect that the inclusion of matter fields in the theory will not produce major changes to the general procedure.

\bigskip

\noindent
{\it Note added}:\\
In Ref.~\cite{Krugz} a careful regularization of both methods (A) and (B) has been carried out. Out of this analysis new contributions have been found for both methods bringing them into agreement.

\bigskip

\acknowledgments{
We acknowledge discussions with D. Karabali and V.P. Nair. 
This work was partially supported by the Spanish 
grants FPA2010-16963 and FPA2011-25948, and by the Catalan grant SGR2009-00894.
}

\end{document}